\documentclass[traditabstract]{aa}
\usepackage{color}
\usepackage{graphicx}
\usepackage{amsmath}
\usepackage{amssymb}
\usepackage{natbib}
\usepackage{dcolumn}

\bibpunct{(}{)}{;}{a}{}{,} 

\newcommand{\msun}{\,{\rm M}_{\odot}}

\newcommand{\ergs}{\,{\rm erg\,s}^{-1}}
\newcommand{\esc}{\,{\rm erg\,s^{-1}\,cm^{-2}}}
\newcommand{\kms}{\,{\rm km\,s^{-1}}}      
\newcommand{\mum}{\,\mu{\rm m}}
\newcommand{\mbh}{$M_{\rm BH}$}
\newcommand{\mmbh}{M_{\rm BH}}
\newcommand{\lbol}{$L_{\rm bol}$}
\newcommand{\mlbol}{L_{\rm bol}}

\begin{document}

\authorrunning{Valencia-S. et al.}
\titlerunning{Is IRAS~01072+4954 a True-Seyfert~2?}
\title{Is IRAS~01072+4954 a True-Seyfert~2?}
\subtitle{Hints from near-infrared integral field spectroscopy}

\author{
       M. Valencia-S.$^{1,2}$
       \and J. Zuther$^{1}$
       \and A. Eckart$^{1,2}$
       \and M. Garc\'{\i}a-Mar\'{\i}n$^{1}$
	   \and C. Iserlohe$^{1}$
       \and G. Wright$^{3}$
}

\offprints{M. Valencia-S. (mvalencias@ph1.uni-koeln.de)}

   \institute{I.Physikalisches Institut, Universit\"at zu K\"oln,
              Z\"ulpicher Str.77, 50937 K\"oln, Germany
         \and
             Max-Planck-Institut f\"ur Radioastronomie,
             Auf dem H\"ugel 69, 53121 Bonn, Germany
		 \and
  	         Astronomy Technology Centre, Royal Observatory Edinburgh, 
		     Edinburgh EH9 3HJ, United Kingdom
              }

\date{Received  / Accepted }

\abstract{
In contrast to the predictions of the unified model, some X-ray unobscured Seyfert~2 galaxies have been discovered in the last decade.
One of them, the starburst/Seyfert composite galaxy IRAS~01072+4954 ($z=0.0236$), has a typical Type-1 X-ray emission, while its optical spectrum resembles an \ion{H}{ii} galaxy and lacks the expected broad lines. 
We performed near-infrared integral-field observations of this object with the aim to determine the nature of its nuclear emission and to find indications for the existence or absence of a broad-line region. 
Several reasons have been proposed to explain this peculiar emission.  
We studied the validity of these hypotheses, including the possibility for this galaxy to be a True-Seyfert~2. We found little obscuration toward the nucleus $A_V = 2.5\,{\rm mag}$, and a nuclear star-formation rate $\Sigma_{\mathrm{{\it SFR}}} < 11.6 \msun\,{\rm yr^{-1} kpc^{-2}}$, which is below the average in Seyferts. Unresolved hot-dust emission with $T \sim 1150\,{\rm K}$ seems to indicate the presence of a torus with its axis close to the line of sight. 
We found that IRAS~01072+4954 hosts a low-mass black hole with an estimated mass of $\mmbh \sim 10^5\msun$. Its bolometric luminosity is $\mlbol \simeq 2.5 \times 10^{42}\ergs$, which yields a high accretion rate with an Eddington ratio $\lambda_{\rm Edd} \simeq 0.2$.
If the relations found in more massive systems also apply to this case, then IRAS~01072+4954 should show broad emission lines with a ${\rm FWHM_{broad}}\sim(400-600)\kms$. Indeed, some indications for such narrow broad-line components are seen in our data, but the evidence is not yet conclusive. This source therefore does not seem to be a True-Seyfert~2, but an extreme case of a narrow-line Seyfert~1, which, due to the faintness of the active nucleus, does not have strong \ion{Fe}{ii} emission in the optical.
}


   \titlerunning{Is IRAS~01072+4954 a True-Sy2?}
   \authorrunning{Valencia-S et al.}
   \maketitle
%

\section{Introduction}
\label{section:Intro}

The unified model of active galactic nuclei (AGN) has been very successful in organizing and integrating a huge range of multiwavelength AGN phenomenology in one simple scheme \citep[e.g.,][]{Barthel89,Antonucci93,Urry95}. The detection of polarized broad lines (BLs) and high X-ray column densities~($N_{\mathrm{H}}>10^{22}$cm$^{-2}$) in most ($\lesssim$50\% and $\sim$96\%, respectively) Seyfert 2 galaxies~(Sy2) supports the idea of the presence of toroidal obscuring material~(`torus') that coves the central engine and the broad-line region~(BLR; \citealp[e.g.,][]{Miller90, Goodrich94}). However, there is an increasing number of observational and theoretical indications that this could be an incomplete picture \citep[see e.g.,][]{Lawrence87,Dopita97}. Sources lacking a BLR, the so-called True-Sy2s, have been suggested  \citep{Nicastro00, Laor03, Elitzur09} and several candidates have been observed \citep{Boisson86, Tran01, PanessaBassani02, Hawkins04, Bianchi08, Shi10, Tran11}. The absence of BLs in their spectra cannot be explained by obscuration given the very low $N_{\mathrm{H}}$ derived from their X-ray emission. Owing to the compelling evidence, extensions of the unified model have emerged \citep{Maiolino95, Dopita97, Elitzur06, Zhang06, Trump11}, but there are still many open questions. Are all AGN the same and differ only by the external obscuring structures? Do they all have a torus? In which ranges of black hole mass and bolometric luminosity is an AGN  able to sustain a BLR?

Several theoretical studies have predicted the disappearance of the BLR in low accretion rate AGN. \citet{Laor03} has suggested, based on the correlation between the radius of the BLR and the bolometric luminosity $\mlbol$ \citep{Kaspi00}, that in low-luminosity objects the BLR shrinks up to a point at which BLs cannot be formed. \citet{Elitzur06} suggested that the BLR and torus are two parts of the same disk-driven wind, which turns from hot and ionized to clumpy and dusty while receding from the central source. In low accretion rate sources ($\mlbol < 10^{42}\ergs$) the outflow is no  longer supported, which causes the disappearance of the torus and, at somewhat lower accretion rates, the  vanishing of the BLR. In a similar approach, \citet{Nicastro00} proposed the BLR-wind to be maintained by the inner part of the accretion disk, which is radiation-pressure dominated. At some critical accretion rate, the transition radius between the radiation-pressure to the gas-pressure dominated disk approaches the innermost marginally stable orbit, and then the BLR cannot form. In addition, it has been suggested that the accretion flow in low-luminosity AGN (LLAGN) can be radiatively inefficient \citep[e.g.][]{Narayan02, Macheto07,Ho09, Trump11}. Models predict that when the Eddington ratio drops below $\sim 10^{-2}$, the accretion disk truncates and a geometrically-thick, optically-thin disk emerges in the inner region, leaving the thin Shakura-Sunyaev disk intact at larger radii. Such sources would have weak emission lines and lack the big blue bump in the optical/UV.

\object{IRAS~01072+4954} is part of a small group of starburst/Seyfert composite galaxies identified by \citet{Moran96} when studying the optical properties of sources that are bright in the far-infrared (IRAS) and that are also detected by ROSAT in X-rays.  The starburst/Seyfert composites were described as objects with optical spectra dominated by starburst features and X-ray emission typical of broad-line AGN. In the Baldwin-Phillips-Terlevich (BPT) diagnostic diagram ([\ion{O}{iii}]/H$\beta$~vs.~[\ion{N}{ii}]/H$\alpha$ emission-line ratios diagram; \citealp{BPT81,VO87}) they are located in between the \ion{H}{ii} galaxies and Seyferts. The active nucleus is visible in the optical through the broader width of the [\ion{O}{iii}] $\lambda\lambda 4959\,,5007$ lines compared to any other permitted or forbidden line. IRAS 01072+4954 does not show the broad H$\alpha$ component identified in other composite galaxies (where the width ranges between 2500 to $3600\kms$) and given the faintness of the narrow H$\alpha$ emission it can be classified as a LLAGN \citep{Ho97}. Using BeppoSAX and Chandra, \citet{Pa05} confirmed the Type-1 character of the spectrum of this source: steep power-law photon-spectral index ($\Gamma=2.1$) and very low hydrogen column density ($N_H < 0.04 \times 10^{22}\,{\rm cm}^{-2}$). Long- and short-term X-ray flux variations were also detected.

The lack of BLs in starburst/Seyfert composite sources has received different explanations: 
(i) The nuclear star formation outshines the optical signatures of accretion (aperture effect; \citealp [e.g.][]{Moran96, Gliozzi10}). 
(ii) Very strong obscuration toward the nucleus hides the BLR and creates an X-ray reflected spectrum that mimics the spectral profile of sources with little absorption (as in the case of the starburst/Seyfert composite NGC 6221; \citealp{Levenson01}). 
 (iii) A clumpy ionized absorber selectively obscures the optical emission while leaving the X-rays unobscured \citep{Georgantopoulos00, Maiolino01}. 
 (iv) The accretion process is radiatively inefficient or the AGN feeding material is blown-away by winds or outflows (this is a common property of LLAGN; \citealp{Macheto07, Ho03}. 
 (v) The AGN is intrinsically weak or there is no BLR at all, so that the source can be considered as a True-Sy2 (like in other unabsorbed Sy2s; \citealp{PanessaBassani02, Panessa09}). High angular resolution data are necessary to provide better constraints on the physical processes at $\sim100\,{\rm pc}$ scales.

Is IRAS~01072+4954 a typical LLAGN or does it host a True-Sy2 nucleus? Why have no broad lines been detected? Given that infrared radiation can arise from deeply embedded -- optically absorbed -- sources, we have carried out high angular and spectral resolution near-infrared (NIR) observations of IRAS~01072+4954, probing a similar region as the previous optical and X-ray data. 
In the present paper we focus on the central $r \approx 75\,{\rm pc}$ emission to uncover the nature of the AGN and offer a possible explanation for the mixed properties in the observed spectral energy distribution. In a follow-up paper \citep{me2012b} we will characterize the surrounding $r \sim 1$\,kpc emission and present a detailed description of the observations and data reduction -- which, for completeness, we briefly summarize here.

The observations and data reduction are described in Sect.\,\ref{sec:obs}. In Sect.\,\ref{sec:results} we study the three main  possibilities for the non-detection of BLs: extinction by cold dust, out-shining nuclear star formation, and obscuration by a hot dust torus. In Sect.\,\ref{sec:bhmabl} the black hole mass and the bolometric luminosity are estimated. The properties of the presumed BLR are derived in Sect.\,\ref{sec:blr}. In Sect.\,\ref{sec:nat} we combine our results with the X-ray and optical information to discuss the nature of the AGN in this source. Summary and conclusions are presented in Sect.\,\ref{sec:fin}. Throughout the study, we use a standard cosmological model with current density parameters $\Omega_{\rm m} = 0.3$, $\Omega_{\Lambda} =0.7$, $H_0=72\kms$.


\section{Observations and data reduction}
\label{sec:obs}

\begin{figure} 
\centering
\resizebox{\hsize}{!}{\includegraphics{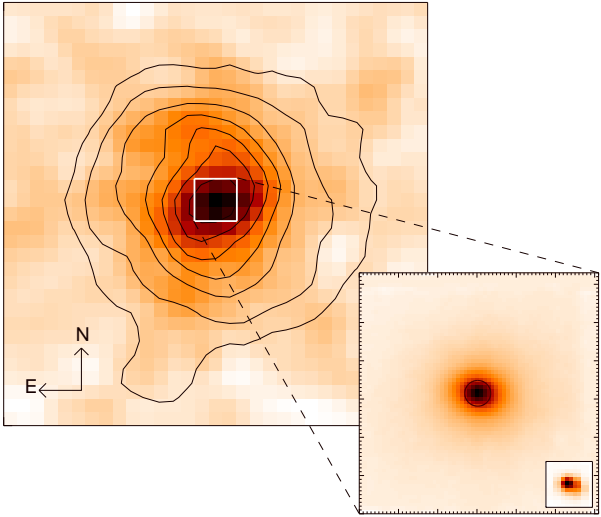}}
\caption{\small
$30'' \times 30''$ 2MASS K-band image of the galaxy IRAS 01072+4954. The overlaid contours correspond to the optical POSS-II image of the galaxy. The levels are 0.3,0.4,0.5,0.6,0.7,0.8 and 0.9 of the optical peak flux. The central thick square represents the $3'' \times 3''$ NIFS FOV. The NIFS K-band continuum is shown in the amplified region. The tick marks on the borders indicate the pixel scale $\sim 0.04''$. The image of the PSF reference star is shown at the bottom-right corner. For this study only the emission from the region enclosed by the circle at the center was considered. 
}
\label{fig:fov}
\end{figure}

The observations of IRAS 01072+4954 were carried out on October 6, 2008 with the Near-Infrared Integral Field Spectrometer NIFS \citep{nifs03} mounted on the 8m ``Frederick C. Gillett'' Gemini North telescope on Mauna Kea, Hawaii. The adaptive optics module ALTAIR was used in Laser Guide Star mode. The $3''\times 3\arcsec$ NIFS field-of-view (FOV) encloses the optical\footnote{Optical image was taken from the Second Palomar Observatory Sky Survey (POSS-II)}/NIR bulge of the galaxy (Fig.\,\ref{fig:fov}). The instrument provides high spatial and spectral resolution simultaneously. 

\noindent

\begin{figure*}
\centering
\includegraphics[width=\textwidth ]{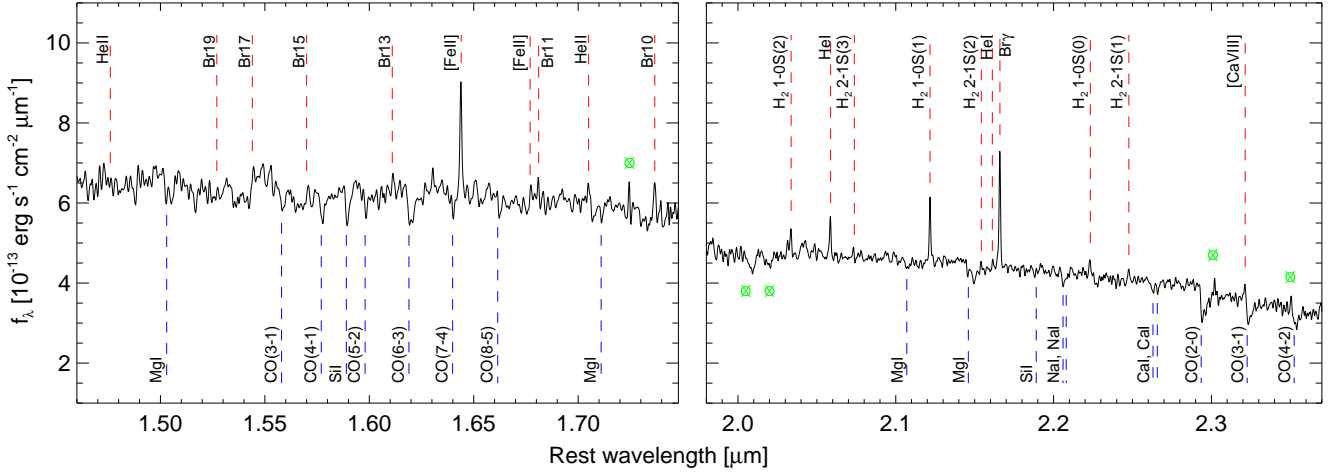}
\caption{\small
H- and K-band spectra of the central $r=0.16\arcsec$. The spectral positions of some emission and absorption lines are signaled. The marked lines do not imply detections. The symbol $\bigotimes$ indicates strong emission/absorption atmospheric features that remained after the reduction process.
}
\label{fig:spec}
\end{figure*}

The observations cover the H- and K-bands, centered on $1.65\mum$ and $2.20\mum$, respectively, with a spectral resolution of $57\kms$. The total integration time was 40~min on source per band. The good atmospheric conditions in Mauna Kea were stable. The point spread function (PSF) was described by a two-dimensional Gauss function fitted to the image of the reference star, \object{J01100963+5010180}. The derived spatial resolution (FWHM of the Gaussian) was $0.17\arcsec$ in H-band and $0.15\arcsec$ in K-band, which correspond to 78\,pc and 72\,pc, respectively. In Fig.\,\ref{fig:fov}, the amplified region corresponds to the NIFS K-band continuum. The image of the PSF reference star is also shown. The circle represents the aperture, with approximately the size of the K-band PSF, used when studying the nuclear emission in K-band. Whenever the combination of H- and K-bands was required, the K-band data were convolved to the resolution of the H-band observations and the integrated emission over a region with the size of the H-band PSF was considered.    

The reduction procedure was performed using the GEMINI IRAF\footnote{IRAF is distributed by the National Optical Astronomy Observatory, which is operated by the Association of Universities for Research in Astronomy (AURA) under cooperative agreement with the National Science Foundation.} package (released Version 2.14, of September 15, 2008).  The NIFS and GNIRS packages were used in combination with generic IRAF tasks. The final reconstruction of the data cubes, flux calibration, and additional data manipulation were accomplished using our own IDL routines. The absolute flux calibration was carried out using the H- and K-band flux densities of Vega ($\alpha$Lyr) computed by \citet{TokunagaandVacca05} for the Mauna Kea NIR filter set. We estimated the uncertainty in our calibration to be $\sim10$\%. Finally, the spectral range of the data cubes was transformed to the rest frame of the galaxy using the published value for the redshift $z=0.023616$. Galactic extinction correction of $E(B-V)=0.156$\,mag \citep{Schlegel98} was applied to the data.


\section{Spectral analysis of the nucleus: Extinction, star formation and dust emission}
\label{sec:results}

Here we describe very briefly the spectrum integrated over the central $r\approx0.16\arcsec$, which corresponds to $r\approx75\,{\rm pc}$ on source. We  find no indications of any classical broad emission at the position of the Br$\gamma$ line. In the subsections we  evaluate the impact of the extinction by cold dust, the nuclear star-formation, and hot dust obscuration (from the torus) on the detection of BL in this source.

\smallskip

The high angular and spectral resolution of NIFS allowed us to resolve the central hundred parsecs of the composite galaxy IRAS~01072+4954. 
Fig.\ref{fig:spec} shows the spectrum extracted at the center from an aperture with the size of the PSF, $r\approx75\,{\rm pc}$. The most prominent emission lines in the H- and K-bands are [\ion{Fe}{ii}] $\lambda1.644\mum$, Br10 $\lambda1.737\mum$, \ion{He}{i} $\lambda2.059\mum$, H$_{2}$(1-0)S(1) $\lambda2.122\mum$ and ~Br$\gamma$ $\lambda2.166\mum$.  Deep absorption lines are also present in the central spectrum.  Several CO absorption bands -- like CO(4-1) $\lambda1.578\mum$,  (6-3) $\lambda1.618\mum$, (7-4) $\lambda1.640\mum$, (2-0) $\lambda2.294\mum$, (3-1) $\lambda2.323\mum$ , and (4-2) $\lambda2.352\mum$ -- as well as \ion{Si}{i} $\lambda1.598\mum$, \ion{Na}{i} $\lambda\lambda 2.206\,,2.209\mum$ and \ion{Ca}{i} $\lambda\lambda 2.263\,,2.266\mum$ can be identified. 
The presence of the AGN could only be inferred from the faint and unresolved emission of the coronal line [\ion{Ca}{viii}] $\lambda2.322\mum$.

\subsection{Active nucleus and broad emission lines}

\begin{figure} 
\centering
\resizebox{\hsize}{!}{\includegraphics{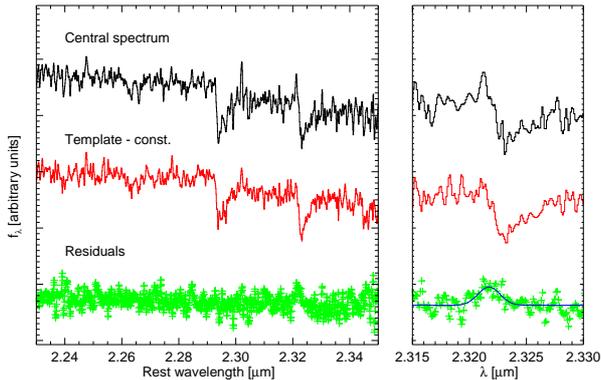}}
\caption{\small
Coronal [\ion{Ca}{viii}] emission in the central spectrum. The spectrum on top is extracted from the center with an aperture of the size of the PSF. The spectrum in the middle is the constructed stellar template, which is shifted vertically for clarity. The residuals from the subtraction of the latter from the former are shown with crosses. In the right panel we show a zoom-in around the [\ion{Ca}{viii}] spectral position and the fit of the line.
}
\label{fig:ca8}
\end{figure}

Coronal lines are reliable AGN tracers, because pure starbursts produce few ionizing photons with energies higher than 54\,eV. With the high spatial resolution data of NIFS, we were able to detect the [\ion{Ca}{viii}] line, which requires an ionizing potential of 127\,eV to be excited. Despite its faintness, the [\ion{Ca}{viii}] line can be seen in the spectrum integrated over an aperture with the size of the PSF. Given the spectral position of this line, just at the border of the \ion{CO}(3-1) band-head, a clear detection is not possible without removing the stellar contribution. Unfortunately, the procedure used to remove the stellar features -- described in Sect.\,\ref{subsec:ccfit} -- also introduces additional noise in the residuals. After applying this procedure the flux of the line, estimated from a Gaussian fit, was $7.9 \times 10^{-17}\esc$, the ${\rm FWHM}\simeq 290\kms$ and the signal-to-noise $(S/N) \simeq 2.5$ (Fig.\,\ref{fig:ca8}).
Although that NIR observations are less affected by dust obscuration and even though NIFS mapped the galaxy nucleus at high angular resolution, a classical broad component with a ${\rm FWHM_{broad}}\ga 2\,000\kms$ was not detected in the spectrum. The median width of the strong emission lines is ${\rm FWHM}\sim 80\kms$.  The criteria for the detection, at three times the level of the noise $\sigma$, of the broad component of a line at wavelength $\lambda$ can be written as
 \begin{equation}
\frac{ F_{{\rm broad,}\lambda}} {3 \sigma \sqrt{ {\rm FWHM}^2_{{\rm broad,}\lambda} - \Delta \lambda^2}} > 1\;,
    \label{eq:detec}
\end{equation}
where $\Delta \lambda$ is the spectral resolution of the instrument and $F_{{\rm broad,}\lambda}$ is the flux of the broad line, which is assumed to be a Gaussian with a full-width-half-maximum of ${\rm FWHM}_{{\rm broad,}\lambda}$.  Accordingly, the minimum flux that the broad-Br$\gamma$ component should have to be detected in our data is $F_{{\rm broad,Br}\gamma}= 1.8\sqrt{{\rm FWHM}^2_{{\rm broad,Br}\gamma} -(56.7)^2}\,\times10^{-19}\esc$ (with ${\rm FWHM}_{{\rm broad,Br}\gamma}$ measured in $\kms$). Assuming a ${\rm FWHM}_{{\rm broad,}\lambda}=2200\kms$  -- which is the median value for the sample of \citet{Ho97width} -- for the BL, Eq.\,(\ref{eq:detec}) implies that a Br$\gamma$ broad component would be detectable if it were about two times brighter than the observed narrow one (as a reference, the flux of the reddening-corrected Br$\gamma$ line is $\simeq 2.1 \times 10^{-16}\esc$).


\subsection{Extinction}
\label{subsec:ext}

The obscuration of the BLR can be caused by galactic structures present on all scales from kiloparsecs down to parsecs \citep[see e.g.,][and references therein]{Bianchi12}. It has been shown, for example, that the orientation of a galaxy has an impact on the classification of its nuclear activity \citep{Shen10,Lagos11}. IRAS~01072+4954 is seen approximately face-on and no signatures of dust lanes crossing the line of sight to the nucleus are visible. Here we measure the extinction\footnote{We refer to extinction $A_{\lambda}$ as the net effect of the absorption and the scattering of the light along the line of sight produced by dust grains of different sizes and chemical compositions \citep[see e.g.,][]{Natta84, Calzetti94}. Therefore, we assume it to be produced by cold dust.} at the center using recombination line ratios and compare it with those from other Sy1 and Sy2 galaxies to estimate the influence of cold dust on the observability of BL.   

\smallskip

To estimate the extinction, it is necessary to make several assumptions about the distribution of the dust, the origin of the emission and the reddening curve \citep[e.g.,][]{RiekeyLebofsky85, Cardelli89, Calzetti00}.  As a first approximation, we assumed a homogeneous scatter-free dust screen placed in front of the nucleus. The fluxes of Br$\gamma$ and Br10 nebular hydrogen-emission lines were used as a probe for extinction according to the expression
\begin{equation}
    A_{V}=50.275\,\times \log \left[ \frac{ \left( F_{\mathrm{Br}\gamma} / F_{\mathrm{Br10}} \right)_{\mathrm{observ}}}{3.025} \right],
    \label{eq:av}
\end{equation}
where $A_V$ is the extinction in the visual band in magnitudes. The theoretically expected line ratio was calculated assuming the case-B recombination for a region with electron density of $10^4\,{\rm cm^{-3}}$ and temperature $T=10^4\,{\rm K}$ \citep{Osterbrock89}. From this expression, we calculated $A_V=2.5\,{\rm mag}$ in the central $r=0.16\arcsec$ spectrum.

The measured $A_V$ is consistent with the median value $\langle A_V \rangle=1.64\,{\rm mag}$ of 27 Sy2 galaxies from the database of \citet{Ho97}, as pointed out by \citet{Rhee05}. In contrast, the median extinction of the 9 Sy1 sources of the same sample was $<0.03\,{\rm mag}$. However, it is still not clear whether this difference between Sy1s and Sy2s is directly related to the obscuration of the BLR by cold dust, because in their study the extinction was determined from the optical Balmer decrement in apertures of $2\arcsec \times4.1\arcsec$, which map areas from a few pc$^2$ (in NGC~3031) to nearly 1\,kpc$^2$ (in NGC~1275). From a detailed comparison of the two types of Seyferts of the same dataset, \citet{Ho03} found differences in the electron densities of the emitting regions and in the environment of the host galaxies. 

To obtain a better spatial correspondence to the region we are considering here ($r\approx 75\,{\rm pc}$), we selected from the literature sources observed with high angular resolution and with extinction measured in the NIR. They are listed in Table\,\ref{tab:syav}, including the aperture within which the fluxes were extracted, translated into a physical scale at the galaxy redshift.   Table\,\ref{tab:syav} shows that (i) The $A_V$ calculated from NIR lines is higher than the median extinction calculated from the optical data. This is expected because NIR observations probe regions with higher optical depths. (ii) The difference in extinction between the two types of objects is less evident or not present at all when measured in the NIR. For example, we found $\langle A_{V,\mathrm{Sy1.5-1}} \rangle =(4.1 \pm 3.0)\,{\rm mag}$, $\langle A_{V,\mathrm{Sy1.8-2}} \rangle =(4.0 \pm 3.1)\,{\rm mag}$. It is known that several Sy2 sources show broad components only in the NIR (e.g., NGC~5506 see \citealp{Nagar02}; NGC~2992 see \citealp{Reunanen03};  Mrk~573 see \citealp{Ramosalmeida08}; \citealp{Veilleux97}). An explanation of this result is not evident, though probably related to the whole obscuring structure around the BLR, including the BL-emitting clouds shadowing themselves, the inclination and composition of the torus and the larger-scale components of the galaxies. Although the central $A_V$ of IRAS 01072+4954 is comparatively low, it is not possible to judge the impact of that extinction on the observability of broad components based only on the extinction value.
\begin{table}
\centering
\caption{NIR extinction measurements from high spatial resolution observations of randomly selected Seyfert galaxies.}
\label{tab:syav}
\centering 
\begin{tabular}{l l c l c}
\hline\hline 
Object & Type & Aperture & $A_V$ & Ref.\\
\hline
NGC~5506 & Sy1\tablefootmark{a} &  93$\times$155\,pc$^2$ & 5.0 & 1\\
NGC~7582 & Sy1\tablefootmark{a} &  200$\times$200\,pc$^2$ &  1.86 & 2 \\
NGC~4151 & Sy1.5 &  20$\times$20\,pc$^2$ &  1.55 & 3 \\
NGC~3783 & Sy1.5 &  $r=$170\,pc &  $\sim$8.0 & 4 \\
Mrk~609  & Sy1.8 &  $r=$160\,pc &  1.26 & 5 \\%
Mrk~1066 & Sy2   &  72$\times$72\,pc$^2$ &  1.84\tablefootmark{b} & 6 \\
ESO~428-G14 & Sy2 &  $r=$85\,pc &  $<$3.0\tablefootmark{c} & 4 \\
Mrk~1157 & Sy2\tablefootmark{d} & 145$\times$145\,pc$^2$ &  5.03 & 7 \\
NGC~1068 & Sy2 & $r=$120\,pc &  3.50 & 8 \\
NGC~5135 & Sy2 & $r=$180\,pc &  9.71 & 9 \\%
\hline
\end{tabular}
\tablefoot{
The $A_V$ was calculated from NIR lines.\\
\tablefoottext{a}{With observed broad Pa$\beta$ in the NIR spectrum.}
\tablefoottext{b}{Calculated from the Pa$\beta/$Br$\gamma$ ratio.}
\tablefoottext{c}{From Fig~1. of \citet{Sch09}.}
\tablefoottext{d}{With observed polarized broad lines.}
}
\tablebib{
(1)~\citet{Nagar02}; (2)~\citet{Rif09}; (3)~\citet{Sch09}; (4)~\citet{Reunanen03}; (5)~\citet{Zuther}; (6)~\citet{Rif10}; (7)~\citet{Rif11}; (8)~\citet{Lucimara10}; (9)~\citet{Bedregal}.
}
\end{table}

\subsection{Star formation}
\label{subsec:sb}

Nuclear star formation can out-shine the BLR signatures. A detailed analysis of star formation and feedback at the nucleus is presented in Valencia-S. et al. (2012b). Here, we estimate its impact on the observed emission from the inner $\approx75$\,pc. We used three different techniques: (i) A NIR diagnostic diagram, which is proposed to distinguish between star formation and AGN as the main ionizing source of the NIR emission lines. (ii) The star-formation rate derived from the Br$\gamma$ luminosity is used as an indicator of the starburst power. (iii) The fraction of the stellar emission in the continuum emission estimated from the equivalent widths of the stellar absorption lines in the spectrum.   

\smallskip

\begin{figure} 
\centering
\resizebox{\hsize}{!}{\includegraphics{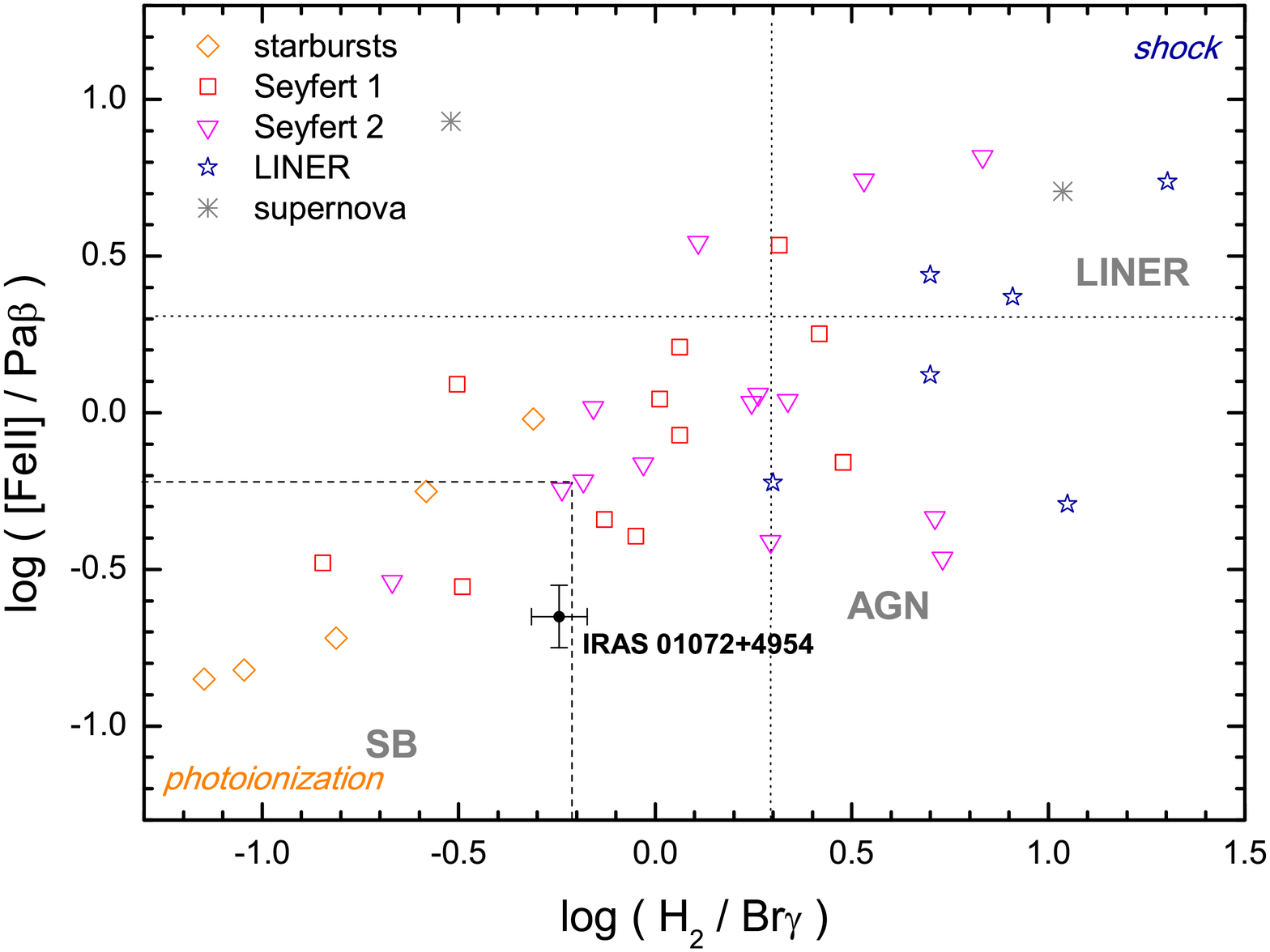}}
\caption{\small
NIR diagnostic diagram. The line ratios obtained from the central region of IRAS~01072+4954 are shown in comparison to the ones calculated from non-spatially resolved observations of other sources taken from the literature. The different type of activity in those objects is represented as shown in the label. The observational division between starbursts, Seyferts and LINERS are shown by dashed lines. The ratio [\ion{Fe}{ii}]$\lambda1.644\mum /{\rm Br}\gamma$ was converted to [\ion{Fe}{ii}]$\lambda1.257\mum /{\rm Pa}\beta$ using theoretical values as explained in the text.
}
\label{fig:nirdiag}
\end{figure}

The broad-band spectral-energy distribution (SED) of IRAS~01072+4954 peaks in the far-infrared ($L(\mathrm{FIR}) = 5.4 \times 10^{43}\ergs$, calculated from the IRAS fluxes following \citealp{Kewley02}).  The colors in that band, $\log(f_{60\mum}/f_{12\mum})>0.52$, $\log(f_{60\mum}/f_{25\mum})>0.52$,  $\log(f_{100\mum}/f_{25\mum})>0.86$, $\log(f_{100\mum}/f_{60\mum})=0.34\pm0.06$, are in the range of those of starburst galaxies 
or active galaxies with higher than 70\% FIR contribution arising from star formation (see e.g., \citealp{Dopita98,Kewley00}).
Although most of the star formation could take place in the disk, high angular resolution is required to isolate it, as much as possible, from the AGN emission. 
To estimate the relative importance of this process at the nucleus, we used the NIR-diagnostic diagram. In analogy to the optical BPT diagram, \citet{Larkin98} and \citet{ardila04, ardila05} suggested that the line ratios of prominent NIR species, [\ion{Fe}{ii}] $\lambda1.257\mum$, Pa$\beta$, H$_2 \lambda2.122\mum$ and Br$\gamma$, could be used to determine the type of nuclear activity. Given that [\ion{Fe}{ii}]$ \lambda1.257\mum$ \,and  Pa$\beta$ were not covered by our observations, we calculated those fluxes from [\ion{Fe}{ii}]$ \lambda1.644\mum$ and Br$\gamma$ using the theoretical value calculated by \citet{nuss88} for the ratio of the iron lines and the case-B for the hydrogen recombination. Previously the spectrum was corrected for extinction using $f_{\lambda,\, \mathrm{corr}}(\lambda) =f_{\lambda,\, \mathrm{observ}}(\lambda) \times 10^{0.4\, A_V\, A_{\lambda,{\rm V}}( \lambda)}$ with $A_V=2.5\,{\rm mag}$ and the parametrization of the interstellar reddening $A_{\lambda,{\rm V}}( \lambda)$ of \citet{Cardelli89}. The originally proposed line ratios are essentially insensitive to obscuration, but we had to account for it because the lines we considered are in different bands. We found $\log({\rm H_2/Br\gamma})= -0.24 \pm 0.07$ and $ \log({\rm [\ion{Fe}{ii}\,1.257]/ Pa\beta})=-0.65 \pm 0.10$. Fig.\,\ref{fig:nirdiag} shows the NIR-diagram where the line ratios of this source are compared with those of the starbursts, Seyferts and LINERs compiled by \citet{ardila05}.  

\noindent

In the diagram, a moderate correlation between both line ratios along different activity types can be recognized. This trend is believed to correspond to an increasing progression from pure photoionization to pure shock excitation of [\ion{Fe}{ii}] and H$_2$. IRAS~01072+4954 falls in between the areas occupied by starbursts and Seyfert galaxies, which implies a mixture of ionization mechanisms of the gas in the central region. 
To better constrain the amount of the stellar contamination at the center, it is necessary to estimate the strength of the emission of young and old stellar populations.   
The latter dominates the NIR emission, while the former oner is relevant at optical wavelengths. However, young populations can also account for $\lesssim 20$\% of the HK-band continuum in starburst galaxies \citep{Rifstarlight}.

Young OB stars emit UV photons that either ionize hydrogen or are absorbed by dust and converted into far-infrared flux. Because the intensity of the hydrogen recombination lines is proportional to the Lyman continuum flux, they are reliable tracers of the star formation in cases where the AGN contribution is negligible. We used the star-formation rate ({\it SFR}) as an indicator of starburst power. Following the calibration of \citet{Panuzzo03}, we calculated it as
 \begin{equation}
\frac{\mathrm{{\it SFR}}}{\msun\,{\rm yr}^{-1}}=\frac{L(\mathrm{Br}\gamma)}{1.585 \times 10^{39}\ergs},
    \label{eq:sfr}
\end{equation}
where $L(\mathrm{Br}\gamma)$ refers to the luminosity of the Br$\gamma$ line. At the center of IRAS~01072+4954, the ${\mathrm{\it SFR}}<0.35\msun\,{\rm yr}^{-1}$, implying a star-formation rate density of $\Sigma_{\mathrm{{\it SFR}}}<11.6\msun \,{\rm yr^{-1}\,kpc^{-2}}$. Here, we did not remove any contribution from the AGN, therefore this value is an upper limit.  To compare with other objects, we calculated the $\Sigma_{\mathrm{{\it SFR}}}$ in the same way for the sample of AGN studied by \citet{Muller11} from the  Br$\gamma$ fluxes and the aperture sizes reported in their Table~2. 
In general, on scales of tens of parsecs, the star formation ranges from $(50-500)\msun \,{\rm yr^{-1}\,kpc^{-2}}$, reaching some $1000\msun \,{\rm yr^{-1}\,kpc^{-2}}$ on parsec scales; over hundreds of parsecs the star-formation rate density reduces to $(1-50)\msun \,{\rm yr^{-1}\,kpc^{-2}}$. A similar result was also found by \citet{Davies07} using a different method for the estimation of the {\it SFR}. Although our reference sample is not statistically significant (11 sources, 5 Sy1s, 4 Sy2s and 2 intermediate types), it is possible to notice that the nucleus of IRAS~01072+4954 is not among the stronger star-forming sources but closer to the lower end of the $ \Sigma_{\mathrm{{\it SFR}}}$ range. In the sample of composite sources studied by \citet{Moran96}, the authors compared the equivalent width of the H$\alpha$ and [\ion{N}{ii}] lines of the composite objects with the spiral and starburst galaxies studied by \citet{Kennicutt92}, arriving at a similar conclusion for the sample: the starburst components in composites are not exceptional among \ion{H}{ii}-galaxies. Therefore, if this source harbors a classical BLR, the star formation seems not to be an obstacle to observe BLs.

While the starburst component at the center is comparatively weak, deep absorption features of an old stellar population can be recognized in the spectrum. An estimate of the fraction of the continuum emission that is produced by intermediate-age and old stars can be obtained from the equivalent width $W$ of the absorption lines. Origlia et al. (1993) and Oliva et al. (1995) showed that the stellar populations of normal galaxies (ellipticals and spirals) span a small range of $W$. In the presence of an AGN, the continuum flux increases due to the emission from hot dust, which has its maximum in the NIR, and as a consequence a decrease in the values of $W$ is observed. In some Seyfert galaxies, a power-law contribution,  originated possibly in the AGN, has also been detected \citep[see e.g.][]{ardila05nlsy, Rif09torus, Rif10}.  Following the same idea, \citet{Davies07} produced synthetic spectra using the code STARS \citep[e.g.,][]{Sternberg98, Thornley00} to reproduce the behavior of the stellar-population properties as a function of time. These authors showed that after 10\,Myr, independent of the star-formation history, the values of the $W_{\ion{CO}(6-3)}$, $W_{\ion{Na}{i}}$ and $W_{\ion{CO}(2-0)}$  vary only by $\sim20$\% around some typical values. Similar results can be obtained with the code SB99 \citep{Leitherer99, Vazquez05}. The equivalent widths predicted by these codes agree with the previous measurements of Origlia et al. (1993) and Oliva et al. (1995). 

\noindent
\begin{figure}
\centering
\resizebox{\hsize}{!}{\includegraphics{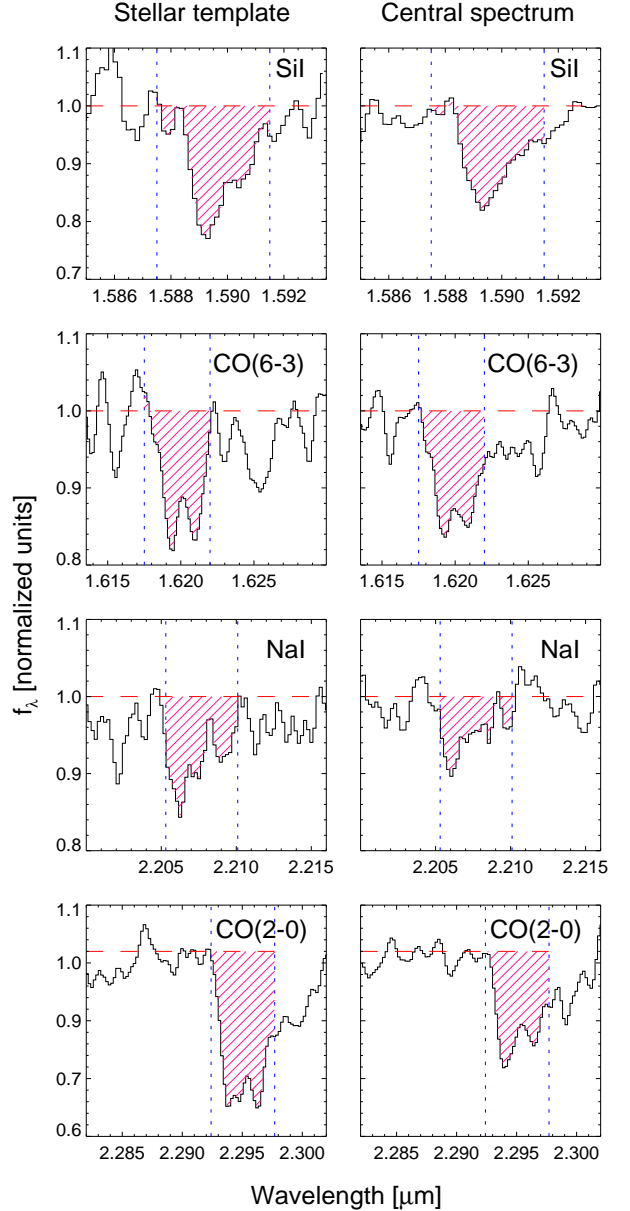}}
\caption{\small
{Absorption lines in the stellar template spectrum (left) and the spectrum extracted at the center (right) of IRAS~01072+4954. The spectra are normalized to the continuum level, which is shown as a dashed line in each case. The vertical dotted lines mark the limits of the integration regions, and shaded areas the equivalent widths.} The measured values are reported in Table\,\ref{tab:ew}. From top to bottom the absorption features are, as indicated, \ion{Si}{i}, \ion{CO}(6-3), \ion{Na}{i} and \ion{CO}{(2-0)}.
}
\label{fig:ews}
\end{figure}

We measured the equivalent widths of the CO(6-3) and CO(2-0) bands, the \ion{Si}{i} and the \ion{Na}{i} lines in the spectrum extracted from the central $r=0.16\arcsec$ and compared them with the theoretically typical ones $W_{\mathrm{int}}$, and with those measured from a stellar template $W_{\mathrm{temp}}$ (Fig.\,\ref{fig:ews}). The template was obtained by integrating the emission over an annulus\footnote{The northern part of the annulus was cut out, because it crosses a region of very recent star formation (see Valencia-S. et al. 2012b)} with inner radius $0.24\arcsec$ and outer radius $0.43\arcsec$. The typical equivalent widths for populations older than 10\,Myr obtained from the STARS and SB99 codes and the measured equivalent widths in the central and in the template spectra are listed in Table\,\ref{tab:ew}. We used the integration limits quoted by \citet{OO93} for the $W_{\mathrm{SiI}}$, $W_{\mathrm{CO(6-3)}}$ and $W_{\mathrm{CO(2-0)}}$ and those by \cite{Forschreiber00} for  $W_{\mathrm{NaI}}$. The stated errors represent $1\sigma$ intervals assuming that the noise over the integrated area is the same as that in neighboring regions selected to contain no obvious stellar-absorption lines. We calculated  the error corresponding to 1\% of uncertainty on the continuum-level estimation to be  $\sim0.4$\,\AA. The fraction of stellar emission in the central aperture $f^{\mathrm{stellar}}$ was calculated as the ratio of $W$ at the center and a reference value, following the prescription of \citet{Davies07}. Taking as reference values the equivalent widths predicted from the synthetic stellar-population codes $W_{\mathrm{int}}$ and the equivalent widths measured in the template spectrum $W_{\mathrm{temp}}$ we obtained two estimates of the stellar fraction in the central region, $f^{\mathrm{stellar}}_{\mathrm{int}}=W / W_{\mathrm{int}}$ and $f^{\mathrm{stellar}}_{\mathrm{temp}}=W / W_{\mathrm{temp}}$ for each absorption feature in the study. These values are also presented in Table\,\ref{tab:ew}.

\begin{table}
\centering
\caption{Equivalent widths measured in the central region $W$, in the stellar template $W_{\mathrm{temp}}$, and predicted by synthetic stellar-population codes $W_{\mathrm{int}}$. The fraction of the continuum at the center that is emitted by stars $f^{\mathrm{stellar}}$ was obtained by comparing $W$ with $W_{\mathrm{temp}}$, and with $W_{\mathrm{int}}$.} 
\label{tab:ew}
\centering 
\begin{tabular}{l l l l l l}
\hline\hline 
Feature & $W$\tablefootmark{a} & $W_{\mathrm{int}}$ & $f^{\mathrm{stellar}}_{\mathrm{int}}$& $W_{\mathrm{temp}}$ & $f^{\mathrm{stellar}}_{\mathrm{temp}}$ \\
 & [\AA ] & [\AA ] & [ \% ] &[\AA ]& [ \% ]\\
\hline
SiI      & $3.5\pm0.2$\tablefootmark{b} & 2.5\tablefootmark{c} & 100\tablefootmark{d} & $4.2\pm0.3$\tablefootmark{b} & 80 \\
CO(6-3)  & $4.9\pm0.3$ & 4.5 &100\tablefootmark{d}  & $4.6\pm0.3$ & 100\tablefootmark{d} \\
NaI      & $3.0\pm0.2$ & 2.5 &100\tablefootmark{d}  & $4.5\pm0.3$ & 65 \\
CO(2-0)  & $8.6\pm0.3$ & 12 &70 &$13.6\pm0.4$ & 65 \\
\hline
\end{tabular}
\tablefoot{ \tablefoottext{a}{From the spectrum integrated over a region of $r\approx75\,{\rm pc}$.}
\tablefoottext{b}{Corrected for the contribution of Br14 $\lambda$1.588.}
\tablefoottext{c}{Calculated using SB99 code.}
\tablefoottext{d}{Set to the maximum value, 100\%.}
}
\end{table}
 
Although the theoretically typical $W_{\mathrm{int}}$ of the \ion{Si}{i} and \ion{Na}{i} are lower than the measured equivalent widths in both absorption features, the values are still consistent within the errors of the measurements and with the scatter of the equivalent widths in stellar populations of different ages. In general, it is possible to notice that the H-band spectrum appears to be completely dominated by the stellar emission, while at least 20\% of the K-band emission has a different origin.  


\subsection{Hot dust emission}
\label{subsec:ccfit}

To investigate the nature of the central non-stellar emission, we subtracted the stellar contribution from the nuclear spectrum and fitted the remaining flux to determine the amount of featureless continuum power-law contribution in these bands and  the temperature of the hot dust. The latter is associated with the presence of the putative torus (or toroidal obscuration) and depends on its orientation with respect to the line of sight toward the observer. Both of these parameters give hints on the viability of observing BLs, if present.

\smallskip

\begin{figure} 
\centering
\resizebox{\hsize}{!}{\includegraphics{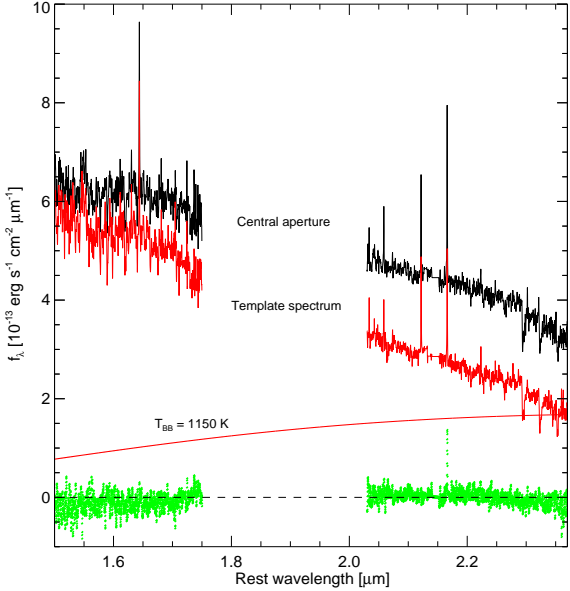}}
\caption{\small Fit of the NIR central spectrum of IRAS 01072+4954. The spectrum of the nucleus integrated over an aperture of $r=0.16\arcsec$ is presented on top. The stellar and hot-dust fitted contributions are shown. They were modeled with the constructed stellar-template (middle spectrum) and a blackbody of $T=1150$\,K, respectively. Small spectral regions with spurious residuals were masked and replaced by the median of the values at their borders. The residuals of the fitting procedure are shown at the bottom with dots. The dashed line marks the zero flux.}
\label{fig:hotdust}
\end{figure}

 Following the method described by \citet{ardila05nlsy}, we assumed the central spectrum to be the linear combination of three components: stellar emission, blackbody radiation from hot dust and a non-thermal source described as a power law $f_{\lambda} \propto \lambda^{\alpha}$. The stellar contribution is represented by the stellar template obtained as explained above (Sect.\,\ref{subsec:sb}) and corrected for extinction. This implies the assumption that the stellar populations at the center and in the surroundings share the same characteristics. The template was scaled to have about the same $W_{\mathrm{CO(2-0)}}$ (within the errors) as the central spectrum and then was subtracted from it. The scaling factor of the template was also restricted such that the emission lines of H$_2$ and [\ion{Fe}{ii}] were not oversubtracted.  The next step was fitting the blackbody and power-law contributions to the residual emission. In the wavelength range of our data, it was not possible to constrain the power-law component and therefore the independence of the blackbody and the power-law contributions could not be assured. In a conservative approach, we fitted the residuals only with a blackbody using the IDL based routine MPFITEXPR \citep{Markwardt09}, which performs a Levenberg-Marquardt least-squares minimization (Fig.\,\ref{fig:hotdust}).  
 To achieve a better approximation to the continuum, small spectral regions with residuals of strong absorption or emission lines were masked.

\noindent

The appropriate estimation of the obscuration in the central spectrum and in the template was critical, because the reddening acts over the spectrum in the opposite way as the hot-dust contribution. Observations over a broader wavelength baseline could help to overcome the degeneracy problem. The central spectrum was corrected for an extinction of $A_V=2.5$\,mag, as discussed in Sect.\,\ref{subsec:ext}. Because the Br10 line was very faint in several regions from which the template was taken, we used the H$-$K color to calculate the extinction acting on the template spectrum $A_{V,\mathrm{temp}}$. We used a $\sim0.5\AA$ window around $1.61\mum$ and $2.20\mum$ to determine the H- and K-band magnitudes, respectively. Assuming that the H$-$K color of a late-type stellar population is about $0.26$\,mag \citep{willner84}, the extinction was computed as
 \begin{equation}
A_{V,\mathrm{temp}}=\frac{({\rm H} - {\rm K})-0.26}{0.0765},
    \label{eq:hmk}
\end{equation}
which in our case implies $A_{V,\mathrm{temp}}=3.8\,{\rm mag}$. The result of the modeling showed that  75\% of the central  $r\approx75\,{\rm pc}$ emission is produced by stars and the remaining 25\% is described by a blackbody of $T \approx 1150\,{\rm K}$ in agreement with the estimates based on the equivalent width of the absorption lines. Fig.\,\ref{fig:hotdust} shows the best fit to the central spectrum. Possible errors in the estimation of $A_{V,\mathrm{temp}}$ affect mainly the value of the temperature of the fitted blackbody. A variation of 0.5\,mag in the $A_{V,\mathrm{temp}}$ causes an increase/decrease of $\sim80\,{\rm K}$. Lower obscuration of the stellar component results in higher temperatures, while an $A_{V,\mathrm{temp}} \geq 5.0\,{\rm mag}$ (which implies $T\leq 980\,{\rm K}$) produces negative residuals in the H-band and a low-quality fit. The weak trend present in the residuals of the best model fit (Fig.\,\ref{fig:hotdust}), though it is consistent with the error of $\sim 10$\% in the flux calibration, it could indicate that the stellar extinction is slightly overestimated -- and consequently result in an underestimation of the temperature.  

Previous studies that used the same method in NIR spectra \citep[e.g.,][]{ardila05, Rif09torus, Rif10} or broad-band SED modeling  \citep[e.g.,][]{Marco98, Kishimoto07, Exposito11} found temperatures of $\gtrsim 1000\,{\rm K}$ in Type-1 AGN and $\lesssim 800\,{\rm K}$ in Type-2s. From JHK spectroscopy, \citet{land11} also found temperatures ranging from 1100 to $\sim1700$\,K for a sample of 23 Type-1 sources. Recently, \citet{Mor11} showed that the NIR emission of $\sim80$\% Type-1 AGN can be explained by emission from hot, pure-graphite dust clouds. 
This hot and unresolved central emission has been interpreted as evidence for the putative torus predicted by the unified model. 
Given that for $T \gtrsim 1000\,{\rm K}$ most astrophysical grain compositions sublimate \citep{Salpeter77, Barvanis87, Granato94}, emission at these temperatures must correspond to dust located very close to the accretion disk (clouds in the inner region of a torus?), while farther out the temperature of the dust decreases rapidly as a result of a larger distance from the source and the shadowing effect caused by other dust clouds \citep{Elitzur08}. A dust temperature of $\approx1150$\,K in IRAS 01072+4954 indicates that we have -- at least partially -- a clear view toward the center. With the wavelength range of our data, it was not possible to fit the expected featureless continuum power-law emission coming from the central source. The presence of this component with a typical flux density $f_{\lambda} \propto \lambda^{-0.5}$ would imply a higher temperature of the hot dust emission. 


\section{The black hole mass and the bolometric luminosity}
\label{sec:bhmabl} 

In this section, we use scaling relations and correlations valid for classical AGN -- with black hole masses between $10^7$ to $10^9\msun$ -- to characterize the active nucleus of this galaxy. First, we fit the brightness profile and stellar absorption features to find the luminosity and stellar velocity dispersion of the bulge. We use both quantities to estimate the black hole mass and briefly discuss the validity of the employed relations. In the second part, we use the hard X-ray luminosity $L_{2-10\mathrm{keV}}$ of IRAS~01072+4954 reported by \citet{Pa05}, to estimate its bolometric luminosity $\mlbol$ and the mass-accretion rate.

\smallskip

\subsection{Brightness profile}
\label{subsec:bp}

\begin{figure} 
\centering
\resizebox{\hsize}{!}{\includegraphics{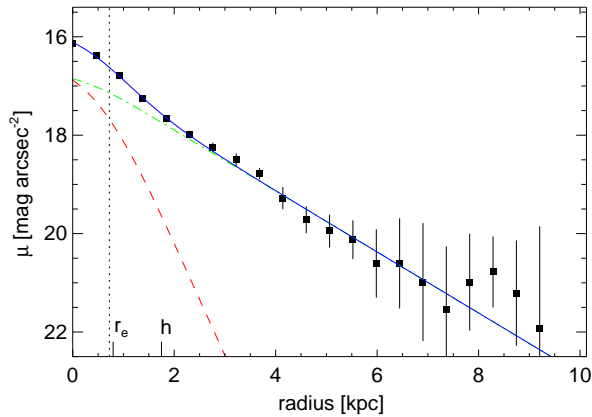}}
\caption{\small K-band brightness profile derived from the 2MASS image. The solid line is the best fit of the data (filled squares). The two components of the model, bulge (dashed line) and disk (dot-dashed line), are shown. In the horizontal axis the positions of the effective radius $r_e$ and the scale radius $h$ are marked. The width of the 2MASS K-band PSF is signaled with the vertical dotted line. The sky level is reached at $\sim21$\,mag\,arcsec$^{-2}.$
}
\label{fig:prof2mass}
\end{figure}

To estimate the black hole mass via scaling relations with the bulge of the host, it is necessary to perform a proper decomposition of the photometric components (AGN, bulge, disk) of the galaxy.  The only available NIR images of the whole galaxy are provided by the 2MASS project data base \citep{2mass}. The faintness of the source in the 2MASS observations does not allow a two-dimensional decomposition. As a first approximation we obtained the brightness profile of the galaxy $I_{\rm observ}(r)$ integrating the 2MASS K-band image over circular rings of $1 \arcsec$ width, centered at the continuum peak. The 2MASS K-band point-spread-function $\mathrm{PSF}_{\mathrm{2MASS}}$ was derived from a star in the field and modeled as a Gaussian with ${\rm FWHM=3.34 \arcsec}$. The bulge was assumed to follow a general Sersic profile with effective radius $r_e$ and index $n$ as parameters. The disk was described by an exponential profile with scale radius $h$. The fit was performed using the MPFIT routine to minimize the expression $|I_{\rm observ}(r) - I_{\rm model}(r)|^2$, where
\begin{eqnarray}
I_{\rm model}(r) &=&  
  \biggl[ c_0 \, \exp \left\{-\nu_n[(r/r_e)^{1/n} -1] \right\} \nonumber \\
 &+& c_1 \, \mathrm{e}^{(-x/h)} \biggr] \ast \mathrm{PSF}_{\mathrm{2MASS}} , 
    \label{eq:2massfit}
\end{eqnarray}
here, $c_0$ and  $c_1$ are constants, $\nu_n$ is a function of the Sersic index\footnote{$\nu_n$ can be determined from $n$ using the  approximation $\nu_n = 2\,n - 1/3 + 4/(405\,n) + 46/(25515\,n^2) +  \mathcal{O} \left( n^{-3} \right)$.} and `$\ast$' means convolution. We set $n=1$ as the minimum value for the Sersic index. 
Fig.\,\ref{fig:prof2mass} presents the best fit to the 2MASS K-band data, with the photometric parameters $r_e=(0.8 \pm 0.3)\,{\rm kpc}$, $n\approx1$ and $h=(1.75 \pm 0.8)\,{\rm kpc}$. As Fig.\,\ref{fig:prof2mass} shows, the fitted effective radius, which corresponds to $\sim 1.7 \arcsec$, is about the same as the ${\rm PSF_{2MASS}}$, meaning that probably the bulge component is not resolved in the 2MASS image.
\noindent

\begin{figure} 
\centering
\resizebox{\hsize}{!}{\includegraphics{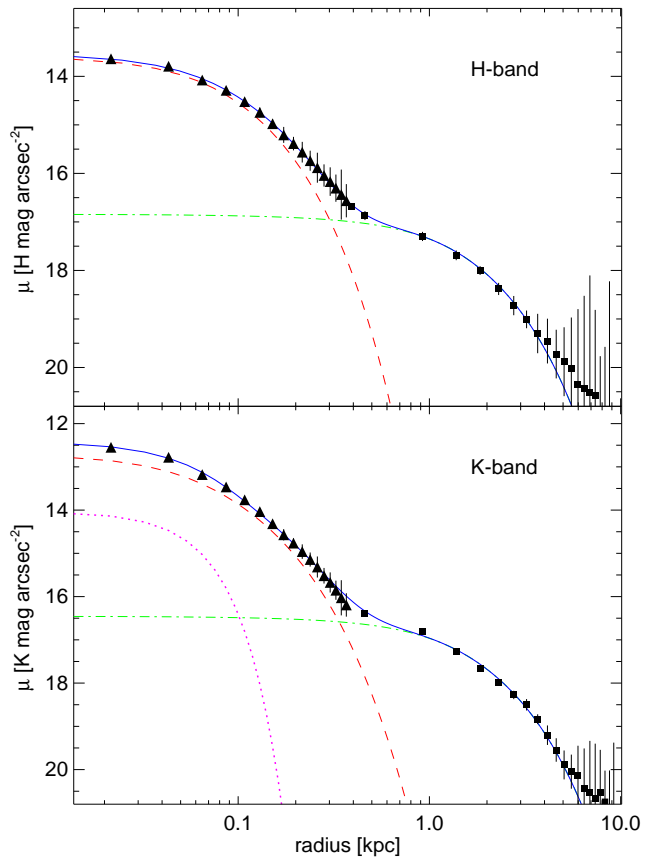}}
\caption{\small Brightness profile of IRAS~01072+4954 obtained by combining the NIFS (triangles) and 2MASS (squares) data. The H-band profile is shown in the upper panel and K-band at the bottom. In both cases, the dashed line corresponds to a Sersic profile fitted to the bulge, and the dash-dotted line to an exponential function that fits the disk component. The solid lines show the overall fits. The dotted line in the lower panel represents the AGN + hot dust contribution fixed to the 25\% of the inner $r\approx 75\,{\rm pc}$ emission,  based on the results presented in Sect.\,\ref{subsec:ccfit}.   The photometric parameters of the profiles in both bands are listed in Table\,\ref{tab:HKsp}. 
}
\label{fig:prof}
\end{figure}

A second attempt to obtain the bulge parameters was perfomed including the NIFS data. Fig.\,\ref{fig:prof}  shows the best fit to the H- and K-band NIFS+2MASS data. In the case, the Sersic profile was convolved to the resolution of the NIFS observations and the exponential function to that of the 2MASS image. One extra component was used to model the K-band profile, where the AGN + hot dust contribution was modeled as a Gaussian with the width of the NIFS point-spread-function $\mathrm{PSF}_{\mathrm{NIFS}}$. Hence, the surface-profile model that we used was 
\begin{eqnarray} 
\label{eq:profboth}
I_{\rm model}(r) &=& k_0 \, \mathrm{PSF}_{\mathrm{NIFS}}\nonumber\\
&+& \left[ k_1 \, \exp \left\{-\nu_n[(x/r_e)^{1/n} -1] \right\} \right] \ast \mathrm{PSF}_{\mathrm{NIFS}}   \nonumber\\
&+&  \left[ k_2 \, \mathrm{e}^{(-x/h)} \right] \ast \mathrm{PSF}_{\mathrm{2MASS}}  ,
\end{eqnarray}
where $k_0, k_1$ and $k_2$ are coefficients and the other symbols have the same meaning as before. In the H-band, the contribution from the AGN and the hot dust are negligible -- or not significant compared to the stellar one -- , therefore when fitting the H-band profile, we used $k_0 =0$. On the other hand, for the K-band modeling, $k_0$ was fixed to account for the detected 25\% of non-stellar emission in the central $r=0.16''$ (see Sect.\,\ref{subsec:ccfit}).  The photometric parameters $r_e$, $n$ and $h$ as well as the surface brightness at the center $I_0$ obtained from the H- and K-band fits are reported in Table\,\ref{tab:HKsp}. 

\noindent

\begin{table}
\centering
\caption{Results from NIFS+2MASS surface brightness profile fits. Photometric parameters derived from the fitting procedure are listed. Black hole mass estimates in both bands were obtained using the bulge luminosity vs. $\mmbh$ relation.}
\label{tab:HKsp}
\centering 
\begin{tabular}{l l l}
\hline\hline 
  & H-band & K-band \\
\hline
 $n$         &  $1.02 \pm 0.07$ & $1.17 \pm 0.08$\\
 $r_e$ ~[pc] & $152.5 \pm 12.9$ &  $159.2 \pm 17.8$\\
 $h\,$  ~[kpc]   & $1.42 \pm 0.26$ & $1.47 \pm 0.53$\\
 \hline
 $I_0$\tablefootmark{a} ~[$10^3$~L$_{\odot,\lambda}$ pc$^{-2}$]  & $34.2 \pm 0.5 $ & $74.6 \pm 2.0$ \\
 $L_{\mathrm{\lambda,bulge}}$\tablefootmark{a} ~[$10^9$~L$_{\odot,\lambda}$] & $1.71 \pm 0.60 $ & $3.24 \pm 1.91$ \\
 \mbh \tablefootmark{a,b} ~[$10^6\msun$ ] & $ 1.2^{+1.8}_{-0.7} $ & $2.5^{+4.7}_{-1.7}$\\
\hline
\end{tabular}
\tablefoot{ \tablefoottext{a}{Upper limit, as explained in the text.}
\tablefoottext{b}{Calculated using \citet[][their Table~2]{Marconihunt03}.}
}
\end{table}
Brightness profiles similar to those shown in Fig.\,\ref{fig:prof} have been observed in pseudobulge galaxies. \citet{KormendyKen04} presented some examples in which a combination of HST WFPC2 and 2MASS JHK photometry allowed them to study the photometric components of nearby galaxies that have pseudobulges in great detail. However, our fit has to be taken as a first approximation to the actual brightness profile, because the data sets do not properly overlap. The continumm maps made from the NIFS data cubes extend until $\sim 1.3\arcsec$ ($\sim 0.6\,{\rm kpc}$) where the noise level is reached, which corresponds to $\lesssim \mathrm{PSF}_{\mathrm{2MASS}}$. While it is clear that the NIFS observations captured the bulge of the galaxy and that the 2MASS image describes the extended component better, 
 physically both components overlap along the projected radius, implying that in Eq.\,(\ref{eq:profboth}) the second and third terms are not independent of each other, as assumed. Given the very different resolutions and sensitivities, it is possible that the flux at the center of the 2MASS image is spread out and therefore part of the disk flux at small radii is missing.   
On the other hand, in \citet{me2012a} we showed that IRAS~01072+4954 is a barred galaxy. The bar component does not seem to dominate in the central $\sim1$\,kpc, but its contribution was not taken into account when constructing the 1D profile. As an overall result, the bulge luminosity could be overestimated by up to one order of magnitude. A proper decomposition of the surface brightness requires high angular resolution observations of the whole galaxy or a major portion of it.


\subsection{Black hole mass estimation}
\label{subsec:Mbh}

The black hole mass has found to be correlated with some properties of the host galaxy bulge \citep[e.g.,][]{Magorrian99, Gebhardt00, Ferrarese00, Marconihunt03}. Here we use the velocity dispersion of the stars in the bulge and its NIR luminosity to estimate $\mmbh$, under the assumption that the scaling relations involving these quantities are valid in this case.

The specific H- and K-band bulge luminosities $L_{\mathrm{H,bulge}}$ and $L_{\mathrm{K,bulge}}$ were obtained by integrating the Sersic profiles that best fitted the H- and K-band bulge surface brightness. The values of $L_{\mathrm{H,bulge}}$ and $L_{\mathrm{K,bulge}}$, which are given in Table\,\ref{tab:HKsp}, are of the order of few $10^9$ solar luminosities. From the scaling relations between $\mmbh$ - $L_{\mathrm{H,bulge}}$ and $\mmbh$ - $L_{\mathrm{K,bulge}}$ found by \citet[][see their Table~2]{Marconihunt03}, we estimated the black hole mass to be $\sim (1-2) \times 10^6\msun$. The $\mmbh$ obtained in this way are presented also in Table\,\ref{tab:HKsp}. The errors do not include the intrinsic rms scatter of the correlations of $\sim0.52$\,dex. 

\begin{figure*} 
\sidecaption
 \includegraphics[width=12cm]{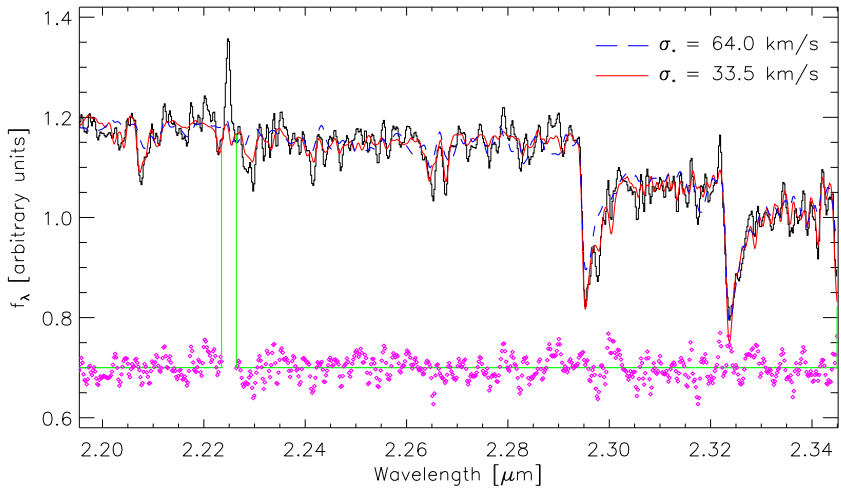}
   \caption{\small Fit of the stellar kinematics of the K-band spectrum integrated over an aperture $r=r_e=160\,{\rm pc}$. The observed spectrum is shown as a thin solid line. Two different fits to the data are shown: the one obtained when using the whole stellar NIFS library (blue dashed line) and the one with the selected NIFS stellar templates (red solid line).  Small spectral regions with emission or spurious lines were masked for the fit and in the plot are delimited  by the vertical lines. The residuals, shifted vertically for clarity, are shown at the bottom with dots. 
}
\label{fig:coppxf}
\end{figure*}

Another estimate of the black hole mass can be obtained using the relation between the $\mmbh$ and the stellar velocity dispersion $\sigma_{\ast}$. To measure $\sigma_{\ast}$, we fitted stellar templates to the K-band spectrum integrated over an aperture of $r\approx0.3\arcsec$, which corresponds to the $r_e$ found above. We used the penalized pixel fitting routine (pPXF) of \citet{Cappellari04}, which minimizes the difference between the galaxy and convolved stellar templates. The algorithm  allows one to apply different weights to the templates, and returns the radial velocity $V$, the velocity dispersion $\sigma_{\ast}$, and higher-order Gauss-Hermite moments $h_3$ and $h_4$ of the line of sight velocity distribution (LOSVD). However, the routine is sensitive to template mismatch and therefore one has to select the templates to closely match the galaxy spectrum. In our case, the stellar templates were taken from the Gemini spectroscopic library of NIR  stars observed with the NIFS IFU in K-band, because of the need for high spectral resolution \citep{Winge09}. The library contains 11 giant and supergiant stars with stellar types from G8 to M5.  When using the full library as an input of the pPXF routine, the quality of the fit was poor and the velocity dispersion was clearly overestimated (Fig.\,\ref{fig:coppxf}). Therefore, we selected a subsample of objects whose absorption features better resemble the H- and K-band continuum of the galaxy.  
 For this selection we used the medium-resolution ($R\sim2000 - 3000$) NIR library of \citet{Ivanov04}, given the lack of high-resolution  H-band templates. We compared the HK  spectra of all stars with the same spectral type as those of the Gemini library with the galaxy-continuum spectrum and discarded the objects with stellar type earlier than K2. The subsample consisted of stellar spectra of the following types: K2III, K5Ib, K5II, K5III, M2III, M3III and M5III.  Spectral regions with emission lines and residuals of sky lines were masked. Fig.\,\ref{fig:coppxf} shows the best fit of the LOSVD with the parameters $V=(-17.1\pm 1.6)\kms$, $\sigma_{\ast}=(33.5 \pm 3.7)\kms$, $h_3=0.059\pm0.046$ and $h_4=0.011\pm0.076$. 
The error was estimated via Monte Carlo simulations where the pPXF routine was applied to 1000 realizations of an input spectrum. That spectrum was created adding Poisson noise to a stellar template to the level of the data i.e., $S/N\sim 40$. Then, the effects of undersampling (because the velocity dispersion is comparable to the velocity scale) were simulated allowing the velocity scale ($\kms$ per pixel) of the input spectrum to change in such a way that the second moment of the LOSVD varied in the range $\sim 14 - 140\kms$.

\noindent

Assuming that AGN follow the same $\mmbh$ - $\sigma_{\ast}$ correlation found for inactive galaxies, we used the \citet{Tremaine02} relation to estimate the black hole mass with the correlation parameters found by \citet[][their Eq.~3.]{Gultein09} to obtain $\mmbh=(0.68^{+0.97}_{-0.40}) \times 10^5\msun$ \footnote{Using the updated values of \citet[][their Table~2.]{Grahammbh} the black hole mass reduces to $\mmbh=(3.41^{+6.18}_{-2.20}) \times 10^3\msun$}. The intrinsic rms scatter of the correlation is $\sim0.4$\,dex. Several authors have pointed out that the $\mmbh$ - $\sigma_{\ast}$ relation in AGN could have a different slope \citep{GreenHo06, Woo10, Xiao11}. With the \citet{Xiao11} correlation ($\sim0.46$\,dex) we obtained  $\mmbh =(1.43^{+1.13}_{-0.64}) \times 10^5\msun$, which agrees with the previous findings.

The estimates of the $\mmbh$ from both scaling relations differ by about one order of magnitude. This can be reconciled given the scatter of the correlations and the approximations introduced when obtaining the HK bulge luminosities. But also, as pointed out by A. Graham (private communication), the $\mmbh$ - $L_{\mathrm{bulge}}$ relation has been calibrated using mainly bright (elliptical) galaxies and probably, it is not suitable in the case of low-luminosity ($M_{B, {\rm bulge}} \gtrsim-20.5\,{\rm mag}$) and low black hole mass ($\mmbh \lesssim 7\times 10^7 \msun$) sources. Elliptical galaxies show a bimodality in the photometric parameters that depends on the luminosity and might extend to disk galaxy bulges \citep[see e.g.,][]{Kormendy09}. This dichotomy seems to be related with the merger history of the galaxies -- i.e., bright core galaxies are a result of dry merger events, while fainter spheroids might form via wet merging -- and could imply different $\mmbh$ - $L_{\mathrm{bulge}}$ relations for both kind of systems \citep{Grahammvl}. For low-luminosity objects, it could be steeper, following a similar behavior as the $\mmbh$ vs. spheroid dynamical mass relation \citep[][his Fig.~1]{Grahammvl}. Therefore, in our particular case, we considered the black hole mass obtained in this way as a conservative upper limit.

On the other hand, pseudobulges and barred galaxies do not seem to follow the same scaling relations defined by ellipticals and classical ($n\sim4$) bulges. Pseudobulges appear to lie below both correlations (e.g., \citealp{Jiang11,KormendyNat}, but see also \citealp{Graham11}), while barred galaxies have a larger scatter around them \citep{Hu08,Graham08,Gultein09,Grahammbh} or follow a steeper relation \citep{Xiao11}. It is not clear whether IRAS~01072+4954 has a pseudobulge. The low Sersic index of the profile is an indication of it, but it is not sufficient evidence and the stellar kinematics do not offer conclusive results. The presence of a bar-like structure is revealed by an increment up to $\sim 50\kms$ of the stellar velocity dispersion at ${\rm P.A.}\sim100^{\circ}$, everywhere else in the central $r<500$\,pc, it remains constant at $\sim 35\kms$  (Valencia-S. et al. 2012a), consistent with the velocity dispersion found here for the bulge.


 \subsection{Bolometric luminosity}
 \label{subsec:lbol}

In contrast to other methods, fitting the broad-band spectrum of the source can provide quite an accurate determination of \lbol. However, given the lack of a nuclear SED, we applied a bolometric correction to the hard X-ray luminosity $L_{2-10\mathrm{keV}}$. \citet{Marconi04} and \citet{Hopkins07} have found relations for the bolometric correction as a function of luminosity based on calculated and observed QSO-SEDs, respectively. Both functions gave consistent results, $\log(\mlbol/\ergs)=42.45$ within the $\sim0.1$ scatter of the expressions. However, \citet{vasudevan07} have shown that for narrow-line Seyfert~1 galaxies (NLSy1s) bolometric luminosities derived through bolometric corrections can differ significantly from those obtained via SED-fitting. Given the similarities of IRAS~01072+4954 with the NLSy1s, it is important to look at the possible deviations introduced by the application of that correction derived for a typical AGN population. In the sample of NLSy1s studied by \citet{vasudevan07}, six out of nine sources have fitted-SED bolometric luminosities higher (two sources the same within the errors and one lower) than those estimated via the bolometric correction. Hence, in NLSy1s the $L_{\mathrm{bol}}$ obtained via bolometric correction might be underestimated, and in our source it can be considered as a possible lower limit. This means that  $L_{\mathrm{bol}} \gtrsim 2.5 \times 10^{42}\ergs$ and that the central massive black hole in IRAS 01072+4954 is accreting at a high rate $\lambda_{\mathrm{Edd}} \sim 0.2$ (assuming $\mmbh\sim10^5\msun$). Such a surprising result for a LLAGN is obtained because although its luminosity is low (typical for a LLAGN), its black hole mass is very low compared with classical AGN. The Eddington ratio obtained in this way agrees with the expected value of $\lambda_{\mathrm{Edd}}=0.30 \pm 0.05$ from the $\lambda_{\mathrm{Edd}}$ - $\Gamma$ relation found by \citet{Shemmer08} in unabsorbed, luminous radio-quiet AGN. Here, we used $\Gamma= 2.1$, which is the photon index of the power-law fitted to the X-ray spectrum \citep{Pa05}. The agreement between the values of the Eddington ratio found by these completely independent methods supports to the assumption that the black hole mass is of the order of $10^5\msun$.

\begin{figure} 
\centering
\resizebox{\hsize}{!}{\includegraphics{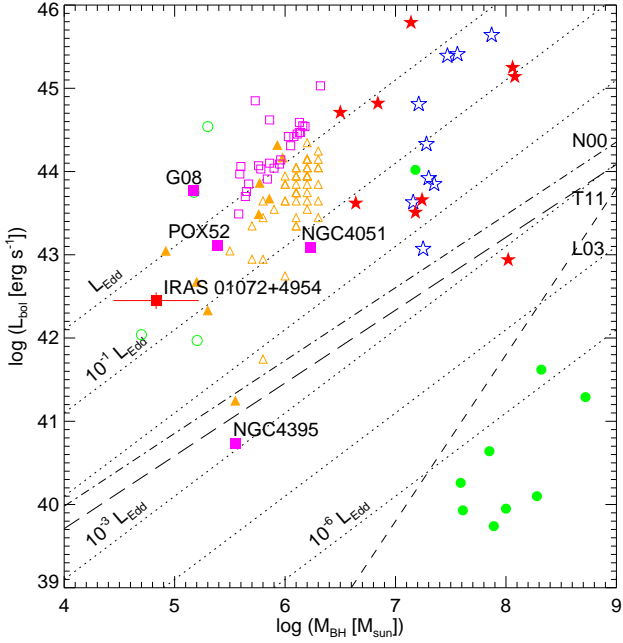}}
\caption{\small Black hole mass vs. bolometric luminosity diagram. The dotted lines  correspond to different Eddington ratios. Theoretical upper limits of True-Sy2s predicted by different authors are also marked: \citet{Nicastro00} `N00' (dot-dashed line), \citet{Laor03} `L03' (short-dashed line) and \citet{Trump11} `T11' (long-dashed line). The location of IRAS~01072+4954 is compared with other sources from the literature. Sy1 and Sy2 from \citet{Singh11} are symbolized as filled-red and empty-blue stars, respectively;  IMBH compiled by \citet{Dewangan08} and by \citet{GreeneHoimbh} are solid and empty orange triangles;
 True-Sy2 candidates of \cite{Laor03} and low-$\mmbh$ Sy2s from \citet{Thornton09} are filled and empty green circles, respectively. NLSy1s with ${\rm FWHM_{broad,H\alpha}} \leq 700\kms$ from the sample of \citet{Zhou06} are plotted with empty magenta squares. Other well-known NLSy1s: NGC~4395, NGC~4051, and SDSS~J114008.71+030711.4 indicated as `G08', and the dwarf-elliptical Sy1 POX~52 are marked with filled magenta squares. 
}
\label{fig:laor}
\end{figure}

The estimated black hole mass and bolometric luminosity of IRAS~01072+4954 place it in the \mbh -$L_{\mathrm{bol}}$ diagram far from the region where theoretically the True-Sy2 are located (Fig.\,\ref{fig:laor}). Instead, it is placed in the area occupied by intermediate mass black holes (IMBHs) and well-studied NLSy1s. The diagram also shows the limits  predicted by different authors for the disappearance of the BLR. For example, in the model proposed by \cite{Laor03}, this could happen at ${\rm FWHM_{BLs}} > 25\,000\kms$. In Fig.\,\ref{fig:laor} the dashed line, labeled L03, represents ${\rm FWHM_{BLs}} = 25\,000\kms$.

 \noindent
 
\section{The broad-line region in IRAS 01072+4954}
\label{sec:blr}  
 
In this section we use the derived black hole mass and bolometric luminosity to deduce the properties of the expected BLR.  We show some indications of the presence of the predicted BL emission in our data. We also study other possibilities for the origin of the emission feature observed at the position of Br$\gamma$.

\smallskip

Does IRAS 01072+4954 have a BLR?  The origin of the broad-line region is still not known. Theoretical works relate it with some sort of wind/outflow from the accretion disk \citep[e.g.,][]{Murray95,  Elitzur06, Nicastro00, Elitzur09, Czerny11}. Models associate the production of this wind with the bolometric luminosity of the accretion disk \lbol, the mass of the black hole $\mmbh$ and the accretion rate $\dot{m}$.\footnote{Here, $\dot{m}$ is the dimensionless accretion rate, defined as $\dot{m} \equiv \dot{M}/ \dot{M}_{\mathrm{Edd}}$, with $\dot{M}_{\mathrm{Edd}}=2.4 \times (\mmbh/10^8)\msun\,{\rm yr^{-1}}$, assuming a radiative efficiency of 0.1 and $\mmbh$ measured in solar masses.}. If $\dot{m} \gtrsim 0.01$, it is possible to equate the accretion rate to the Eddington ratio $\lambda_{\mathrm{Edd}} \equiv \mlbol/L_{\mathrm{Edd}}$. When $\dot{m} < 0.01$ the accreting fluid becomes radiatively inefficient and $\lambda_{\mathrm{Edd}}$ decreases more rapidly than $\dot{m}$. For example, in LLAGN $\langle \lambda_{\mathrm{Edd}} \rangle \lesssim 10^{-3}$ for Sy1s and $\sim 10^{-6}$ for Sy2s, LINERs and transition objects \citep[for a review see e.g.,][]{Ho08}.

\subsection{Properties of the predicted broad-line region}
 \label{subsec:propblr}
 
If IRAS~01072+4954 behaves as other AGN, it must have a BLR. In the following we will assume that the properties of the BLR scale with the luminosity as has been shown for black holes with higher masses \citep[e.g.][]{NetzerLaor93,Kaspi00, Bentz09} to derive the observationally expected fluxes and widths of the BLs. 

\smallskip

From the \citet{Kaspi00} relation, with the updated values of \citet{Bentz09}, we estimated the size of the BLR to be $R_{\mathrm{BLR}}\sim 1$\,light-day, equivalently $\sim10^{-3}\,{\rm pc}$ or $\sim10^{5}$\,R$_{\mathrm{S}}$ (where R$_{\mathrm{S}} \equiv 2GM_{\mathrm{BH}}/c^2$ is the Schwarzschild radius of the black hole). The uncertainty is abound 1 order of magnitude, mainly due to the error of the zero-point of the correlation. This $R_{\mathrm{BLR}}$ is expected from the specific luminosity at 5100\AA, $L_{5100}\approx 1.3 \times 10^{41}\ergs$, which was estimated from the bolometric luminosity.\\ 

In Seyferts and QSO, the strength of the hydrogen recombination lines scales with the X-ray luminosity \citep[e.g.,][]{Kriss80, Ward88}. The total H$\alpha$ flux (narrow+broad components, when present) of LINERS and LLAGN compared with their X-ray fluxes (soft and hard X-rays) seem to follow the same trend \citep[e.g.,][]{Koratkar95, Terashima00, Ho01}, suggesting that they could be scaled-down versions of more massive and powerful AGN.  Recently, \citet{SternLaor2012} studied the relation between the broad H$\alpha$ emission and the luminosity in a sample of about 3600 Type-1 AGN selected from the Sloan Digital Sky Survey, with H$\alpha$ luminosities in the range $7\times10^{39}\ergs$ to $10^{44}\ergs$. They found that the luminosity of the broad H$\alpha$ component is related to the specific luminosity at 2\,keV as
\begin{equation}
\log \left[ \nu L_{\nu}(2\mathrm{keV}) \right] = 0.79 \, \log \left(  L_{\mathrm{broad, H}\alpha} \right) +0.45 \,,
    \label{eq:lbha}
\end{equation}
where both luminosities are given in units of $10^{42}\ergs$. Moreover, assuming that the BLR emitting clouds are virialized (${\rm FWHM^2} \propto \mmbh/R_{\mathrm{BLR}}$; \citealp[e.g.,][]{Laor98, Netzer09}), the authors found that the characteristic parameters of the emitted broad H$\alpha$ line, i.e. ${\rm FWHM}_{{\rm broad,H}\alpha}$ and luminosity, are related to the black hole mass through 
\begin{equation}\begin{split}
\log \left( \frac{ \mmbh}{\msun} \right) = &7.40 + 2.06\, \log \left( \frac{{\rm FWHM}_{{\rm broad,H}\alpha} } {1000\kms} \right) \\
&+ 0.545\, \log \left( \frac{ L_{\mathrm{broad, H}\alpha}}{10^{44} \ergs} \right) \,,
    \label{eq:mha}
\end{split}\end{equation}
with a dispersion of $\sim 0.25$\,dex. Using these expressions, we found that the expected flux of the broad H$\alpha$ line is $F_{\mathrm{broad, H}\alpha}\simeq 4.8 \times10^{-14}\esc$ with a width of $\mathrm{FWHM}_{\mathrm{broad, H}\alpha}\simeq 435\kms$. %
If on the other hand we use  $L_{\mathrm{broad, H}\alpha} \approx \mlbol /130$, which was also  proposed by \citet{SternLaor2012}, then the expected values are $F_{\mathrm{broad, H}\alpha}\simeq 1.7 \times10^{-14}\esc$ and $\mathrm{FWHM}_{\mathrm{broad, H}\alpha} = 600\kms$ \footnote{Using the relations obtained by \citet[][their Eqs.(5-6)]{Xiao11} we obtained $\mathrm{FWHM}_{\mathrm{broad, H}\alpha} \sim 750\kms$.}. 

\begin{figure} 
\centering
\resizebox{\hsize}{!}{\includegraphics{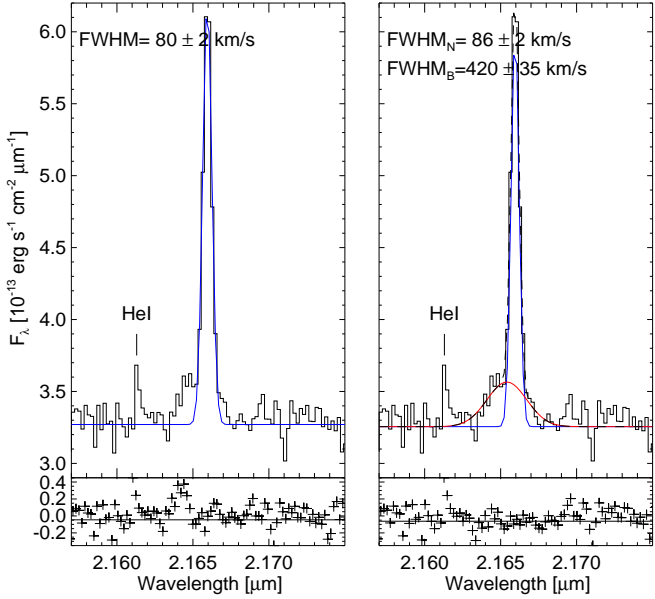}}
\caption{\small Fit of the Br$\gamma$ line emission. A single Gaussian fitted to the data is shown in the left panel. The right panel shows the data fitted with two Gaussians, where the thin solid lines correspond to the individual components and the dashed line to the overall fit. The FWHM of the fitted lines are marked in each case. At the bottom of both panels the residuals of the individual fits are marked with crosses. 
}
\label{fig:brg}
\end{figure}

Whether or not such a component is present in the optical spectrum \citep[see][their Fig.~5]{Moran96} is not clear. Although their spectral resolution ($\sim 200\kms$) is sufficient to measure such a width, the data cover an area of $\sim (2 \times 1)\,{\rm kpc}^2$ on source and the H$\alpha$ line is blended at the base with the neighboring [\ion{N}{ii}] lines. Can we see broad lines with those properties in our observations? In the NIFS data, the strongest recombination line is Br$\gamma$. We assumed $\mathrm{FWHM}_{\mathrm{broad, Br}\gamma} \simeq \mathrm{FWHM}_{\mathrm{broad, H}\alpha}$ and $F_{\mathrm{broad, Br}\gamma}\approx F_{\mathrm{broad, H}\alpha}/100$ to scale the BL properties in the two bands. Note that, formally, the theoretical case-B is not applicable \citep[but see][]{RheeLarkin00, Zhou06, Kim00}; nevertheless, we used it as a first approximation. According to Eq.\,(\ref{eq:detec}) the deteccion of such a broad Br$\gamma$ emission is at about the limit of our instrumental capabilities. We checked in the NIFS K-band central spectrum for indications of a BL of that characteristics. Fig.\,\ref{fig:brg} shows a zoom around $2.166\mum$ where in the left and right panels, one (only narrow) and two (narrow and broad) simple Gaussian components, respectively, were fitted to the data. In the second case, the broad component has a $\mathrm{FWHM}_{\mathrm{broad, Br}\gamma} \simeq 420\kms$ and $F_{\mathrm{broad, Br}\gamma} \simeq 1.0\times10^{-16}\esc$, with a peak shifted  by $0.7\times 10^{-3}\mum$. 

\noindent

To check the possibility of this component being a BL, or the tip of a BL, we performed  several tests.
We fitted the Br$\gamma$ line with a two Gaussian components, to account for the narrow and the broad emissions, in every spatial pixel of the FOV. As a result, the broad component peaks and is wider at the center, while the narrow component peaks a few pixels toward the north, at a position $\sim 100$\,pc away from the center at the redshift of the galaxy. However, that could be a result of the $S/N$, which increases towards the nucleus, so we took increasingly larger apertures and found that the FWHM of the `broad' component decreases systematically with the aperture radius. On the other hand, if the feature observed around $2.166\mum$ originates in the BLR, it should not be present in the profile of forbidden lines like [\ion{Fe}{ii}].  Given that the densities in the BLR are higher than the critical densities required for forbidden lines to form, both kinds of lines originate from spatially separated regions and are, therefore, kinematically de-coupled. 
When comparing the LOSVD of Br$\gamma$ and [\ion{Fe}{ii}] emissions, using e.g. channel maps, we did not find similarities. 
A direct comparison between the Br$\gamma$ and the [\ion{Fe}{ii}] line shapes at the nucleus is not possible, unfortunately. The reason is that the continuum around the spectral position of the [\ion{Fe}{ii}] line $\lambda1.644\mum$ cannot be determined accurately enough to recognize the presence of any faint `broader' emission. The main drawbacks are the lack of spectral templates with high-spectral resolution in the H-band and the presence of strong features in the continuum around $\lambda1.644\mum$, such as the CO(7-4) absorption band starting at $\lambda1.640\mum$ and a residual-sky line at $1.645\mum$. In general, although the fitted flux and width of the `broad' Br$\gamma$ component agree with the predicted ones, this result is not a firm detection ($S/N=2.8$), but it is an interesting coincidence that allows us to speculate about the nature of the observed Br$\gamma$-`hump'. A confirmation of the existence of such faint BLs will require high-angular and high-spectral resolution observations of stronger lines like Pa$\alpha$ or Pa$\beta$.

\subsection{Other explanations for the emission at 2.16\,$\mu$m}

Here we briefly present other possible situations that could create the 'broad' feature at 2.16\,$\mu$m.
\smallskip

The presence of outflows in the NIR spectrum of Seyfert galaxies is traced by the [\ion{Fe}{ii}] emission \citep{Blietz94,Rif11}. In sources like NGC 4151 \citep{Sch09}, Mrk~1066 \citep{Riffel11} and Mrk~1157 \citep{Rif11}, the kinematics of the ionized gas is found to be consistent with previously observed radio emission of jet structures. In Valencia-S. et al. (2012b), we show that in the inner $\sim250\,{\rm pc}$ of IRAS~01072+4954 the LOSVD of the [\ion{Fe}{ii}] emitting gas can be interpreted as an outflow with a line of sight velocity $V_{\mathrm{LOS, [FeII]}}\lesssim -60\kms$. The kinematics of ionized-hydrogen (Br$\gamma$) is consistent with motion in a disk, but extra components cannot be discarded. To consider the possibility that the Br$\gamma$-`hump' corresponds to a blue-shifted emission from wind, we modeled the Br$\gamma$ line with two narrow-Gaussian components, one for the observed line and the other for the blue-shifted wing. Although the quality of the fit is good, the resulting shift of the wing-component is $\sim -180\kms$, which would represent a clear deviation from the disk-profile that is not observed. When mapping the two thin components across the FOV, the velocity field of the blue-shifted component, which represents the Br$\gamma$-`wind', does not follow the LOSVD of the [\ion{Fe}{ii}], which argues against an outflow as the origin of the Br$\gamma$-`hump'.

Another possibility is the presence of a blended emission from other species close to Br$\gamma$. Given that we also detected \ion{He}{i}\,(${\rm 2s^1S-2p^1P^0}$) at $2.0587\mum$, we looked for other helium emission lines within $0.002\mum$ around $2.1661\mum$ (the resolution of the instrument at that wavelength is $\sim 4\times 10^{-4}\mum$). Theoretically, two helium lines could be there: \ion{He}{ii} (14-8) at $2.1653\mum$ and \ion{He}{i}\,(4-7) at $2.1660\mum$. The latter, if present, is impossible to resolve out of Br$\gamma$. Of the former, we calculated the maximum flux contribution based on the upper limit of the flux of the undetected \ion{He}{ii}\,(14-7) at $1.4882\mum$. Using the recombination coefficients of \citet{HummerStorey87} assuming case-B, $T=10^4\,{\rm K}$ and $N_{\mathrm{e}}=10^4\,{\rm cm^{-3}}$, we obtained $F_{\mathrm{HeII(14-8)}}<0.1\times 10^{-16}\esc$. This implies that a potential helium line could at most account for $\sim10$\% of the observed emission in the blue-shifted wing.


\section{On the nature of the source}
\label{sec:nat}

IRAS~01072+4954 is a composite galaxy that has low X-ray absorption toward the center while its optical spectrum lacks broad emission lines. This source might harbor a True-Sy2 nucleus, hence the main objective of this study was to investigate the nature of its peculiar emission. 

Our NIR observations revealed low obscuration in the inner 75\,pc and the presence of dust with $T>1000$\,K, which indicates a clear view toward the nucleus, and is consistent with the X-ray data. 
If the idea that the toroidal obscuration and the BLR are connected products of the same disk-driven wind is true \citep[e.g.][]{Elitzur08}, then this source must have a BLR given the clear signature of hot dust in the central HK-spectrum. Another argument against the True-Sy\,2 hypothesis is the high Eddington ratio ($\lambda_{\mathrm{Edd}} \sim 0.2$), as it is shown in Fig.\,\ref{fig:laor}. This corresponds to a high/soft-accretion state with a geometrically thin optically thick disk. 

Assuming the validity of the \mbh - $L_{\mathrm{bulge}}$ and \mbh - $\sigma_{\ast}$ relations, IRAS~01072+4954 hosts a black hole of $\sim 10^5\msun$ and it shares the same X-ray properties as other IMBHs: short-term variability and $\Gamma$ $\sim1.7-2.6$ \citep[e.g.,][]{GreeneHoXrimbh,Dewangan08,Desroches09}\footnote{Here we refer to the photon index obtained when a single power-law is fitted to the 0.3-10\,keV data, as used by \citet{Dewangan08}. Other authors describe the X-ray spectra using two power-law functions, hence derive a soft- and a hard- photon index, $\Gamma_{\mathrm{S}}$ and $\Gamma_{\mathrm{H}}$. Then using $\Gamma_{\mathrm{S}}$ to characterize NLSy1 is justified because the emission of a Shakura-Sunyaev disk peaks close to the soft X-ray domain.}.
If the properties of the BLR scale with the luminosity as they do in more massive ($\mmbh \sim 10^{6-8}\msun$) AGN, the width and the flux of the expected broad emission lines in this source are below the lower end of the values observed in classical AGN and in the low range of those in NLSy1s. For example, in the homogeneous sample of $\sim$2000 NLSy1s from the SDSS DR3 by \citet{Zhou06}, there are $\sim$500 objects with $\mathrm{FWHM}_{\mathrm{broad, H}\alpha}\leq 1200\kms$, of which $\sim$25 ($\sim1$\%) are below $700\kms$. Those 25 sources have $\langle L_{\mathrm{broad, H}\alpha} \rangle =4.3 \times 10^{41}\ergs$; in comparison, we estimated $L_{\mathrm{broad, H}\alpha}\sim (2.0-6.0) \times 10^{40}\ergs$ for our source.
IRAS~01072+4954 could be an extreme NLSy1, but it does not show the characteristic  \ion{Fe}{ii} emission in the optical spectrum. \citet{Ai01} suggested the existence of two kinds of IMBH sources, distinguishing the NLSy1s for their well-known features, i.e., strong \ion{Fe}{ii} emission,  significant soft X-ray excess, and high Eddington ratios. The other IMBHs would have flatter spectral slopes (more typical of Sy1s), non-ubiquity of the soft X-ray excess, and a low accretion rate.  The closest and by now the most studied case is NGC~4395 \citep[see   e.g.,][]{Filippenko89, Peterson05, Laor07}. \citet{Ai01} proposed the Eddington ratio to be the driving mechanism for the observed properties. However, the properties of those two classes become mixed in the case of IRAS~01072+4954. \citet{Zhou06} have already commented on the existence of ``\ion{Fe}{ii}-deficient'' NLSy1 in sources with high accretion rate and very narrow broad-Balmer lines. Moreover, if the physical conditions for iron production in IMBHs are the same as in more massive AGN, then, according to the correlations found by \citet{Dong11}, in sources with Eddington ratios of $\gtrsim 0.2$ (as this one), the expected flux of the broad \ion{Fe}{ii} line is about the same as the expected flux of the broad H$\beta$ component, and the flux of the narrow \ion{Fe}{ii} is a tenth of it ($10 \times
F_{\mathrm{narrow,\ion{Fe}{ii}}\lambda4570} \sim F_{\mathrm{broad,\ion{Fe}{ii}}\lambda4570} \sim F_{\mathrm{broad, H}\beta}$). This means that for very faint broad components, even for objects with high accretion rate, strong \ion{Fe}{ii} is not expected to be observed.

IRAS~01072+4954 might have a pseudobulge given its low Sersic index, a common characteristic of objects with IMBHs (a counterexample is Pox~52, a dwarf elliptical galaxy with  $\mmbh \simeq 3\times10^5\msun$; \citealp{Thornton08}), but the photometrical and kinematical evidence is not conclusive. It contains a bar-like structure, which is taken to be indicative of secular evolution \citep[e.g.,][]{KormendyKen04, Mathur11, Xivry11}. High-angular resolution observations of the whole galaxy are required to perform a better photometric decomposition and to establish the presence of nuclear structures related with the black hole feeding and feedback mechanisms.

Here, it is important to point out that the optical, NIR, and X-ray observations were not taken simultaneously. Therefore, intermittent obscuration generated by, e.g., a clumpy torus cannot be discarded as a reason for the absence of BLs \citep[see e.g.,][]{Risaliti05, Longinotti09, Marchese12}. As mentioned before, the source presents long-term variability in the X-rays. For example, from 1990 to 2002 the soft X-ray flux decreased by about 1.1 orders of magnitude, but in 2005 it increased again. These variations could be related to a change in the accretion mode, and that would imply a variation in the bolometric luminosity. However, it is unlikely that a variable X-ray absorber could also cover/uncover significant parts of the BLR and be responsible for the True-Sy2 appearance. Here we always referred to the measurements performed by \citet{Pa05}, assuming that by the time of our observations (2008) the X-ray emission did not change dramatically. In consequence, we expect our main conclusions still to be valid.    


\section{Summary and conclusions}
\label{sec:fin}

We have discussed the nature of the active nucleus in IRAS~01072+4954, an unobscured starburst/Seyfert composite galaxy. The observations were performed with the integral-field spectrometer NIFS operated with the ALTAIR laser-guide adaptive-optics module at the Gemini-North Telescope. We based our analysis on H- and K-band data from the central $r\approx75\,{\rm pc}$ with spectral resolution $\sim 60\kms$. 

IRAS~01072+4954  has an X-ray emission typical of Sy1s ($N_H < 0.04 \times 10^{22}$\,cm$^{-2}$, $\Gamma\sim$2.1 and strong short-term variability), but the optical spectrum, taken from a $\sim(1 \times2)\,{\rm kpc^2}$ region, lacks the expected broad lines. We studied the main hypotheses found in the literature for the apparent absence of the broad components: 

\begin{itemize}

\item Extinction from cold dust along the line of sight: The extinction measured from the hydrogen recombination lines in the NIR is $A_V=2.5$\,mag toward the nucleus, which is consistent with that measured in other Seyfert galaxies from high-angular resolution NIR observations, $\langle A_{V} \rangle \simeq (4 \pm 3)$\,mag, in both types of AGN. We found no relation between the extinction measured in the NIR and the observability of broad lines.
\\
\item Nuclear star formation that outshines the AGN emission: The star formation is a dominant process in this galaxy. The NIR diagnostic diagram places the nuclear line ratios at the border between starburst galaxies and Seyferts. However, in the central region, the star-formation rate density is $\Sigma_{\mathrm{{\it SFR}}}<11.6\msun \,{\rm yr^{-1}\,kpc^{-2}}$, which is below the lower limit of the range $(50-500)\msun\,{\rm yr^{-1}\,kpc^{-2}}$ observed in Seyferts. An older stellar population with ages $\gtrsim 10^{7}\,{\rm yr}$ accounts for about 75\% of the continuum flux.
\\
\item Hot dust obscuration, possibly related to the putative torus of the unified model: Without taking the featureless continuum power-law component into account -- which could not be constrained with our data -- about 25\% of the HK-band continuum is emitted by hot dust at a temperature of $\sim1100$\,K. Considering that in Type-1 sources $T\gtrsim 1000$\,K, whereas in Type-2s $T\lesssim 800$\,K, we interpreted this result as an indication of a clear view toward the center. This is also consistent with the low absorption column density estimated by \citet{Pa05} from X-ray observations.
\\
\item Non-existence of the BLR, hosting a True-Sy2 nucleus: The unresolved hot-dust emission that signifies the presence of a torus is predicted to form and persist while the BLR is launched. Therefore, given the clear signature of hot dust at the center of IRAS~01072+4954, and the high Eddington ratio, the presence of a BLR is expected. 

\end{itemize}

Individually, none of the previous possibilities seem to be a satisfactory explanation. We found that the unusual combination of a low black hole mass, $\mmbh \sim 10^5\msun$, with a low bolometric luminosity, $\mlbol \approx 2.8\times10^{42}\ergs$, implies that the expected BLs are very faint $F_{\mathrm{broad, H}\alpha}\sim (1.7-4.8) \times10^{-14}\esc$ and their widths are of only few hundreds of $\kms$, $\mathrm{FWHM}_{\mathrm{broad, H}\alpha} \sim (400 - 600)\kms$. Thus, we conclude that this is the main reason for the non-detection of the BLs. In the K-band spectrum of the central $r\approx75$\,pc, we observe a `hump' around the position of the Br$\gamma$ line with the theoretically predicted flux and width a broad Br$\gamma$ component should have. A proper detection would require higher $S/N$ data. 

IRAS~01072+4954 is a barred galaxy with an exponential surface brightness bulge profile ($n\sim1.1$ and $r_e=160\,{\rm pc}$) and a stellar-velocity dispersion of $33.5\kms$.  It hosts a low mass black hole that accretes at a rate $\dot{m} \approxeq \lambda_{\mathrm{Edd}} \sim 0.2$. This implies a high/soft accreation state with a geometrically-thin accretion disk. 
These properties are similar to those found in NLSy1s, but IRAS~01072+4954 lacks the \ion{Fe}{ii} emission typical for these sources. High resolution optical spectroscopy of the nucleus is necessary to measure the strength of the \ion{Fe}{ii} and confirm the AGN classification.

%


\vspace{0.5cm}

\begin{acknowledgements}
The authors thank J. Stern, A. Laor and A. Graham for constructive discussions and comments to the paper.  We also thank the anonymous referee for his/her comments, which helped to improve this paper.  
M. Valencia-S. is member of the International Max Planck Research School (IMPRS) for Astronomy and Astrophysics at the MPIfR and the Universities of Bonn and Cologne.
J. Zuther acknowledges support from the European project EuroVO DCA under the Research e-Infrastructures area RI031675.
M. Garc\'{\i}a-Mar\'{\i}n is supported by the German federal department for
education and research (BMBF) under the project numbers: 50OS0502, 50OS0801 \& 50OS1101.
Part of this work was supported by the German Deutsche Forschungsgemeinschaft, DFG, via grant SFB 956 and fruitful discussions with the members of the European Union funded COST Action MP0905: Black Holes in a violent Universe and PECS project No. 98040.
Based on observations  (Program ID: GN-2008B-Q-77) obtained at the Gemini Observatory, which is operated by the Association of Universities for Research in Astronomy, Inc., under a cooperative agreement with the NSF on behalf of the Gemini partnership: the National Science Foundation (United States), the Science and Technology Facilities Council (United Kingdom), the National Research Council (Canada), CONICYT (Chile), the Australian Research Council (Australia), Minist\'erio da Ci${\rm\hat{e}}$ncia e Tecnologia (Brazil) and Ministerio de Ciencia, Tecnolog\'{\i}a e Innovaci\'on Productiva  (Argentina).
\end{acknowledgements}

\bibliographystyle{aa}
\bibliography{bibtexInprep}

\begin{thebibliography}{156}
\expandafter\ifx\csname natexlab\endcsname\relax\def\natexlab#1{#1}\fi

\bibitem[{{Ai} {et~al.}(2011){Ai}, {Yuan}, {Zhou}, {Wang}, \& {Zhang}}]{Ai01}
{Ai}, Y.~L., {Yuan}, W., {Zhou}, H.~Y., {Wang}, T.~G., \& {Zhang}, S.~H. 2011,
  \apj, 727, 31

\bibitem[{{Antonucci}(1993)}]{Antonucci93}
{Antonucci}, R. 1993, \araa, 31, 473

\bibitem[{{Baldwin} {et~al.}(1981){Baldwin}, {Phillips}, \&
  {Terlevich}}]{BPT81}
{Baldwin}, J.~A., {Phillips}, M.~M., \& {Terlevich}, R. 1981, \pasp, 93, 5

\bibitem[{{Barthel}(1989)}]{Barthel89}
{Barthel}, P.~D. 1989, \apj, 336, 606

\bibitem[{{Barvainis}(1987)}]{Barvanis87}
{Barvainis}, R. 1987, \apj, 320, 537

\bibitem[{{Bedregal} {et~al.}(2009){Bedregal}, {Colina}, {Alonso-Herrero}, \&
  {Arribas}}]{Bedregal}
{Bedregal}, A.~G., {Colina}, L., {Alonso-Herrero}, A., \& {Arribas}, S. 2009,
  \apj, 698, 1852

\bibitem[{{Bentz} {et~al.}(2009){Bentz}, {Peterson}, {Netzer}, {Pogge}, \&
  {Vestergaard}}]{Bentz09}
{Bentz}, M.~C., {Peterson}, B.~M., {Netzer}, H., {Pogge}, R.~W., \&
  {Vestergaard}, M. 2009, \apj, 697, 160

\bibitem[{{Bianchi} {et~al.}(2008){Bianchi}, {Corral}, {Panessa}, {Barcons},
  {Matt}, {Bassani}, {Carrera}, \& {Jim{\'e}nez-Bail{\'o}n}}]{Bianchi08}
{Bianchi}, S., {Corral}, A., {Panessa}, F., {et~al.} 2008, \mnras, 385, 195

\bibitem[{{Bianchi} {et~al.}(2012){Bianchi}, {Maiolino}, \&
  {Risaliti}}]{Bianchi12}
{Bianchi}, S., {Maiolino}, R., \& {Risaliti}, G. 2012, AdAst, 2012

\bibitem[{{Blietz} {et~al.}(1994){Blietz}, {Cameron}, {Drapatz}, {Genzel},
  {Krabbe}, {van der Werf}, {Sternberg}, \& {Ward}}]{Blietz94}
{Blietz}, M., {Cameron}, M., {Drapatz}, S., {et~al.} 1994, \apj, 421, 92

\bibitem[{{Boisson} \& {Durret}(1986)}]{Boisson86}
{Boisson}, C. \& {Durret}, F. 1986, \aap, 168, 32

\bibitem[{{Calzetti} {et~al.}(2000){Calzetti}, {Armus}, {Bohlin}, {Kinney},
  {Koornneef}, \& {Storchi-Bergmann}}]{Calzetti00}
{Calzetti}, D., {Armus}, L., {Bohlin}, R.~C., {et~al.} 2000, \apj, 533, 682

\bibitem[{{Calzetti} {et~al.}(1994){Calzetti}, {Kinney}, \&
  {Storchi-Bergmann}}]{Calzetti94}
{Calzetti}, D., {Kinney}, A.~L., \& {Storchi-Bergmann}, T. 1994, \apj, 429, 582

\bibitem[{{Cappellari} \& {Emsellem}(2004)}]{Cappellari04}
{Cappellari}, M. \& {Emsellem}, E. 2004, \pasp, 116, 138

\bibitem[{{Cardelli} {et~al.}(1989){Cardelli}, {Clayton}, \&
  {Mathis}}]{Cardelli89}
{Cardelli}, J.~A., {Clayton}, G.~C., \& {Mathis}, J.~S. 1989, \apj, 345, 245

\bibitem[{{Czerny} \& {Hryniewicz}(2011)}]{Czerny11}
{Czerny}, B. \& {Hryniewicz}, K. 2011, \aap, 525, L8

\bibitem[{{Davies} {et~al.}(2007){Davies}, {Oudmaijer}, \& {Sahu}}]{Davies07}
{Davies}, B., {Oudmaijer}, R.~D., \& {Sahu}, K.~C. 2007, \apj, 671, 2059

\bibitem[{{Desroches} {et~al.}(2009){Desroches}, {Greene}, \&
  {Ho}}]{Desroches09}
{Desroches}, L.-B., {Greene}, J.~E., \& {Ho}, L.~C. 2009, \apj, 698, 1515

\bibitem[{{Dewangan} {et~al.}(2008){Dewangan}, {Mathur}, {Griffiths}, \&
  {Rao}}]{Dewangan08}
{Dewangan}, G.~C., {Mathur}, S., {Griffiths}, R.~E., \& {Rao}, A.~R. 2008,
  \apj, 689, 762

\bibitem[{{Dong} {et~al.}(2011){Dong}, {Wang}, {Ho}, {Wang}, {Fan}, {Wang},
  {Zhou}, \& {Yuan}}]{Dong11}
{Dong}, X.-B., {Wang}, J.-G., {Ho}, L.~C., {et~al.} 2011, \apj, 736, 86

\bibitem[{{Dopita}(1997)}]{Dopita97}
{Dopita}, M. 1997, \pasa, 14, 230

\bibitem[{{Dopita} {et~al.}(1998){Dopita}, {Heisler}, {Lumsden}, \&
  {Bailey}}]{Dopita98}
{Dopita}, M.~A., {Heisler}, C., {Lumsden}, S., \& {Bailey}, J. 1998, \apj, 498,
  570

\bibitem[{{Elitzur}(2008)}]{Elitzur08}
{Elitzur}, M. 2008, \nar, 52, 274

\bibitem[{{Elitzur} \& {Ho}(2009)}]{Elitzur09}
{Elitzur}, M. \& {Ho}, L.~C. 2009, \apjl, 701, L91

\bibitem[{{Elitzur} \& {Shlosman}(2006)}]{Elitzur06}
{Elitzur}, M. \& {Shlosman}, I. 2006, \apjl, 648, L101

\bibitem[{{Exposito} {et~al.}(2011){Exposito}, {Gratadour}, {Cl{\'e}net}, \&
  {Rouan}}]{Exposito11}
{Exposito}, J., {Gratadour}, D., {Cl{\'e}net}, Y., \& {Rouan}, D. 2011, \aap,
  533, A63

\bibitem[{{Ferrarese} \& {Merritt}(2000)}]{Ferrarese00}
{Ferrarese}, L. \& {Merritt}, D. 2000, \apjl, 539, L9

\bibitem[{{Filippenko} \& {Sargent}(1989)}]{Filippenko89}
{Filippenko}, A.~V. \& {Sargent}, W.~L.~W. 1989, \apjl, 342, L11

\bibitem[{{F{\"o}rster Schreiber}(2000)}]{Forschreiber00}
{F{\"o}rster Schreiber}, N.~M. 2000, \aj, 120, 2089

\bibitem[{{Gebhardt} {et~al.}(2000){Gebhardt}, {Bender}, {Bower}, {Dressler},
  {Faber}, {Filippenko}, {Green}, {Grillmair}, {Ho}, {Kormendy}, {Lauer},
  {Magorrian}, {Pinkney}, {Richstone}, \& {Tremaine}}]{Gebhardt00}
{Gebhardt}, K., {Bender}, R., {Bower}, G., {et~al.} 2000, \apjl, 539, L13

\bibitem[{{Georgantopoulos}(2000)}]{Georgantopoulos00}
{Georgantopoulos}, I. 2000, \mnras, 315, 77

\bibitem[{{Gliozzi} {et~al.}(2010){Gliozzi}, {Panessa}, {La Franca}, {Saviane},
  {Monaco}, {Foschini}, {Kedziora-Chudczer}, {Satyapal}, \&
  {Sambruna}}]{Gliozzi10}
{Gliozzi}, M., {Panessa}, F., {La Franca}, F., {et~al.} 2010, \apj, 725, 2071

\bibitem[{{Goodrich} \& {Miller}(1994)}]{Goodrich94}
{Goodrich}, R.~W. \& {Miller}, J.~S. 1994, \apj, 434, 82

\bibitem[{{Graham}(2008)}]{Graham08}
{Graham}, A.~W. 2008, \apj, 680, 143

\bibitem[{{Graham}(2011)}]{Graham11}
{Graham}, A.~W. 2011, arXiv:astro-ph/1103.0525

\bibitem[{{Graham}(2012)}]{Grahammvl}
{Graham}, A.~W. 2012, \apj, 746, 113

\bibitem[{{Graham} {et~al.}(2011){Graham}, {Onken}, {Athanassoula}, \&
  {Combes}}]{Grahammbh}
{Graham}, A.~W., {Onken}, C.~A., {Athanassoula}, E., \& {Combes}, F. 2011,
  \mnras, 412, 2211

\bibitem[{{Granato} \& {Danese}(1994)}]{Granato94}
{Granato}, G.~L. \& {Danese}, L. 1994, \mnras, 268, 235

\bibitem[{{Greene} \& {Ho}(2005)}]{GreenHo06}
{Greene}, J.~E. \& {Ho}, L.~C. 2005, \apj, 627, 721

\bibitem[{{Greene} \& {Ho}(2007{\natexlab{a}})}]{GreeneHoimbh}
{Greene}, J.~E. \& {Ho}, L.~C. 2007{\natexlab{a}}, \apj, 670, 92

\bibitem[{{Greene} \& {Ho}(2007{\natexlab{b}})}]{GreeneHoXrimbh}
{Greene}, J.~E. \& {Ho}, L.~C. 2007{\natexlab{b}}, \apj, 656, 84

\bibitem[{{G{\"u}ltekin} {et~al.}(2009){G{\"u}ltekin}, {Richstone}, {Gebhardt},
  {Lauer}, {Tremaine}, {Aller}, {Bender}, {Dressler}, {Faber}, {Filippenko},
  {Green}, {Ho}, {Kormendy}, {Magorrian}, {Pinkney}, \& {Siopis}}]{Gultein09}
{G{\"u}ltekin}, K., {Richstone}, D.~O., {Gebhardt}, K., {et~al.} 2009, \apj,
  698, 198

\bibitem[{{Hawkins}(2004)}]{Hawkins04}
{Hawkins}, M.~R.~S. 2004, \aap, 424, 519

\bibitem[{{Ho}(2003)}]{Ho03}
{Ho}, L.~C. 2003, in \pasp, Vol. 290, Active Galactic Nuclei: From Central
  Engine to Host Galaxy, ed. {S.~Collin, F.~Combes, \& I.~Shlosman}, 379

\bibitem[{{Ho}(2008)}]{Ho08}
{Ho}, L.~C. 2008, \araa, 46, 475

\bibitem[{{Ho}(2009)}]{Ho09}
{Ho}, L.~C. 2009, \apj, 699, 626

\bibitem[{{Ho} {et~al.}(2001){Ho}, {Feigelson}, {Townsley}, {Sambruna},
  {Garmire}, {Brandt}, {Filippenko}, {Griffiths}, {Ptak}, \& {Sargent}}]{Ho01}
{Ho}, L.~C., {Feigelson}, E.~D., {Townsley}, L.~K., {et~al.} 2001, \apjl, 549,
  L51

\bibitem[{{Ho} {et~al.}(1997{\natexlab{a}}){Ho}, {Filippenko}, \&
  {Sargent}}]{Ho97}
{Ho}, L.~C., {Filippenko}, A.~V., \& {Sargent}, W.~L.~W. 1997{\natexlab{a}},
  \apjs, 112, 315

\bibitem[{{Ho} {et~al.}(1997{\natexlab{b}}){Ho}, {Filippenko}, {Sargent}, \&
  {Peng}}]{Ho97width}
{Ho}, L.~C., {Filippenko}, A.~V., {Sargent}, W.~L.~W., \& {Peng}, C.~Y.
  1997{\natexlab{b}}, \apjs, 112, 391

\bibitem[{{Hopkins} {et~al.}(2007){Hopkins}, {Richards}, \&
  {Hernquist}}]{Hopkins07}
{Hopkins}, P.~F., {Richards}, G.~T., \& {Hernquist}, L. 2007, \apj, 654, 731

\bibitem[{{Hu}(2008)}]{Hu08}
{Hu}, J. 2008, \mnras, 386, 2242

\bibitem[{{Hummer} \& {Storey}(1987)}]{HummerStorey87}
{Hummer}, D.~G. \& {Storey}, P.~J. 1987, \mnras, 224, 801

\bibitem[{{Ivanov} {et~al.}(2004){Ivanov}, {Rieke}, {Engelbracht},
  {Alonso-Herrero}, {Rieke}, \& {Luhman}}]{Ivanov04}
{Ivanov}, V.~D., {Rieke}, M.~J., {Engelbracht}, C.~W., {et~al.} 2004, \apjs,
  151, 387

\bibitem[{{Jiang} {et~al.}(2011){Jiang}, {Greene}, \& {Ho}}]{Jiang11}
{Jiang}, Y.-F., {Greene}, J.~E., \& {Ho}, L.~C. 2011, \apjl, 737, L45

\bibitem[{{Kaspi} {et~al.}(2000){Kaspi}, {Smith}, {Netzer}, {Maoz}, {Jannuzi},
  \& {Giveon}}]{Kaspi00}
{Kaspi}, S., {Smith}, P.~S., {Netzer}, H., {et~al.} 2000, \apj, 533, 631

\bibitem[{{Kennicutt}(1992)}]{Kennicutt92}
{Kennicutt}, Jr., R.~C. 1992, \apj, 388, 310

\bibitem[{{Kewley} {et~al.}(2002){Kewley}, {Geller}, {Jansen}, \&
  {Dopita}}]{Kewley02}
{Kewley}, L.~J., {Geller}, M.~J., {Jansen}, R.~A., \& {Dopita}, M.~A. 2002,
  \aj, 124, 3135

\bibitem[{{Kewley} {et~al.}(2000){Kewley}, {Heisler}, {Dopita}, {Sutherland},
  {Norris}, {Reynolds}, \& {Lumsden}}]{Kewley00}
{Kewley}, L.~J., {Heisler}, C.~A., {Dopita}, M.~A., {et~al.} 2000, \apj, 530,
  704

\bibitem[{{Kim} {et~al.}(2010){Kim}, {Im}, \& {Kim}}]{Kim00}
{Kim}, D., {Im}, M., \& {Kim}, M. 2010, \apj, 724, 386

\bibitem[{{Kishimoto} {et~al.}(2007){Kishimoto}, {H{\"o}nig}, {Beckert}, \&
  {Weigelt}}]{Kishimoto07}
{Kishimoto}, M., {H{\"o}nig}, S.~F., {Beckert}, T., \& {Weigelt}, G. 2007,
  \aap, 476, 713

\bibitem[{{Koratkar} {et~al.}(1995){Koratkar}, {Deustua}, {Heckman},
  {Filippenko}, {Ho}, \& {Rao}}]{Koratkar95}
{Koratkar}, A., {Deustua}, S.~E., {Heckman}, T., {et~al.} 1995, \apj, 440, 132

\bibitem[{{Kormendy} \& {Bender}(2011)}]{KormendyNat}
{Kormendy}, J. \& {Bender}, R. 2011, \nat, 469, 377

\bibitem[{{Kormendy} {et~al.}(2009){Kormendy}, {Fisher}, {Cornell}, \&
  {Bender}}]{Kormendy09}
{Kormendy}, J., {Fisher}, D.~B., {Cornell}, M.~E., \& {Bender}, R. 2009, \apjs,
  182, 216

\bibitem[{{Kormendy} \& {Kennicutt}(2004)}]{KormendyKen04}
{Kormendy}, J. \& {Kennicutt}, Jr., R.~C. 2004, \araa, 42, 603

\bibitem[{{Kriss} {et~al.}(1980){Kriss}, {Canizares}, \& {Ricker}}]{Kriss80}
{Kriss}, G.~A., {Canizares}, C.~R., \& {Ricker}, G.~R. 1980, \apj, 242, 492

\bibitem[{{Lagos} {et~al.}(2011){Lagos}, {Padilla}, {Strauss}, {Cora}, \&
  {Hao}}]{Lagos11}
{Lagos}, C.~D.~P., {Padilla}, N.~D., {Strauss}, M.~A., {Cora}, S.~A., \& {Hao},
  L. 2011, \mnras, 414, 2148

\bibitem[{{Landt} {et~al.}(2011){Landt}, {Elvis}, {Ward}, {Bentz}, {Korista},
  \& {Karovska}}]{land11}
{Landt}, H., {Elvis}, M., {Ward}, M.~J., {et~al.} 2011, \mnras, 414, 218

\bibitem[{{Laor}(1998)}]{Laor98}
{Laor}, A. 1998, \apjl, 505, L83

\bibitem[{{Laor}(2003)}]{Laor03}
{Laor}, A. 2003, \apj, 590, 86

\bibitem[{{Laor}(2007)}]{Laor07}
{Laor}, A. 2007, in \pasp, Vol. 373, The Central Engine of Active Galactic
  Nuclei, ed. {L.~C.~Ho \& J.-W.~Wang}, 384

\bibitem[{{Larkin} {et~al.}(1998){Larkin}, {Armus}, {Knop}, {Soifer}, \&
  {Matthews}}]{Larkin98}
{Larkin}, J.~E., {Armus}, L., {Knop}, R.~A., {Soifer}, B.~T., \& {Matthews}, K.
  1998, \apjs, 114, 59

\bibitem[{{Lawrence}(1987)}]{Lawrence87}
{Lawrence}, A. 1987, \pasp, 99, 309

\bibitem[{{Leitherer} {et~al.}(1999){Leitherer}, {Schaerer}, {Goldader},
  {Gonz{\'a}lez Delgado}, {Robert}, {Kune}, {de Mello}, {Devost}, \&
  {Heckman}}]{Leitherer99}
{Leitherer}, C., {Schaerer}, D., {Goldader}, J.~D., {et~al.} 1999, \apjs, 123,
  3

\bibitem[{{Levenson} {et~al.}(2001){Levenson}, {Cid Fernandes}, {Weaver},
  {Heckman}, \& {Storchi-Bergmann}}]{Levenson01}
{Levenson}, N.~A., {Cid Fernandes}, Jr., R., {Weaver}, K.~A., {Heckman}, T.~M.,
  \& {Storchi-Bergmann}, T. 2001, \apj, 557, 54

\bibitem[{{Longinotti} {et~al.}(2009){Longinotti}, {Bianchi}, {Ballo}, {de La
  Calle}, \& {Guainazzi}}]{Longinotti09}
{Longinotti}, A.~L., {Bianchi}, S., {Ballo}, L., {de La Calle}, I., \&
  {Guainazzi}, M. 2009, \mnras, 394, L1

\bibitem[{{Macchetto} \& {Chiaberge}(2007)}]{Macheto07}
{Macchetto}, F.~D. \& {Chiaberge}, M. 2007, in IAU Symposium, Vol. 238, IAU
  Symposium, ed. {V.~Karas \& G.~Matt}, 273

\bibitem[{{Magorrian} {et~al.}(1998){Magorrian}, {Tremaine}, {Richstone},
  {Bender}, {Bower}, {Dressler}, {Faber}, {Gebhardt}, {Green}, {Grillmair},
  {Kormendy}, \& {Lauer}}]{Magorrian99}
{Magorrian}, J., {Tremaine}, S., {Richstone}, D., {et~al.} 1998, \aj, 115, 2285

\bibitem[{{Maiolino} {et~al.}(2001){Maiolino}, {Marconi}, {Salvati},
  {Risaliti}, {Severgnini}, {Oliva}, {La Franca}, \& {Vanzi}}]{Maiolino01}
{Maiolino}, R., {Marconi}, A., {Salvati}, M., {et~al.} 2001, \aap, 365, 28

\bibitem[{{Maiolino} {et~al.}(1995){Maiolino}, {Ruiz}, {Rieke}, \&
  {Keller}}]{Maiolino95}
{Maiolino}, R., {Ruiz}, M., {Rieke}, G.~H., \& {Keller}, L.~D. 1995, \apj, 446,
  561

\bibitem[{{Marchese} {et~al.}(2012){Marchese}, {Braito}, {Della Ceca},
  {Caccianiga}, \& {Severgnini}}]{Marchese12}
{Marchese}, E., {Braito}, V., {Della Ceca}, R., {Caccianiga}, A., \&
  {Severgnini}, P. 2012, \mnras, 421, 1803

\bibitem[{{Marco} \& {Alloin}(1998)}]{Marco98}
{Marco}, O. \& {Alloin}, D. 1998, \aap, 336, 823

\bibitem[{{Marconi} \& {Hunt}(2003)}]{Marconihunt03}
{Marconi}, A. \& {Hunt}, L.~K. 2003, \apjl, 589, L21

\bibitem[{{Marconi} {et~al.}(2004){Marconi}, {Risaliti}, {Gilli}, {Hunt},
  {Maiolino}, \& {Salvati}}]{Marconi04}
{Marconi}, A., {Risaliti}, G., {Gilli}, R., {et~al.} 2004, \mnras, 351, 169

\bibitem[{{Markwardt}(2009)}]{Markwardt09}
{Markwardt}, C.~B. 2009, in \pasp, Vol. 411, Astronomical Data Analysis
  Software and Systems XVIII, ed. {D.~A.~Bohlender, D.~Durand, \& P.~Dowler},
  251

\bibitem[{{Martins} {et~al.}(2010){Martins}, {Rodr{\'{\i}}guez-Ardila}, {de
  Souza}, \& {Gruenwald}}]{Lucimara10}
{Martins}, L.~P., {Rodr{\'{\i}}guez-Ardila}, A., {de Souza}, R., \&
  {Gruenwald}, R. 2010, \mnras, 406, 2168

\bibitem[{{Mathur} {et~al.}(2012){Mathur}, {Fields}, {Peterson}, \&
  {Grupe}}]{Mathur11}
{Mathur}, S., {Fields}, D., {Peterson}, B.~M., \& {Grupe}, D. 2012, \apj, 754,
  146

\bibitem[{{McGregor} {et~al.}(2003){McGregor}, {Hart}, {Conroy}, {Pfitzner},
  {Bloxham}, {Jones}, {Downing}, {Dawson}, {Young}, {Jarnyk}, \& {Van
  Harmelen}}]{nifs03}
{McGregor}, P.~J., {Hart}, J., {Conroy}, P.~G., {et~al.} 2003, in Society of
  Photo-Optical Instrumentation Engineers (SPIE) Conference Series, Vol. 4841,
  Society of Photo-Optical Instrumentation Engineers (SPIE) Conference Series,
  ed. {M.~Iye \& A.~F.~M.~Moorwood}, 1581

\bibitem[{{Miller} \& {Goodrich}(1990)}]{Miller90}
{Miller}, J.~S. \& {Goodrich}, R.~W. 1990, \apj, 355, 456

\bibitem[{Mor \& Trakhtenbrot(2011)}]{Mor11}
Mor, R. \& Trakhtenbrot, B. 2011, \apjl, 737, L36

\bibitem[{{Moran} {et~al.}(1996){Moran}, {Halpern}, \& {Helfand}}]{Moran96}
{Moran}, E.~C., {Halpern}, J.~P., \& {Helfand}, D.~J. 1996, \apjs, 106, 341

\bibitem[{{M{\"u}ller-S{\'a}nchez} {et~al.}(2011){M{\"u}ller-S{\'a}nchez},
  {Prieto}, {Hicks}, {Vives-Arias}, {Davies}, {Malkan}, {Tacconi}, \&
  {Genzel}}]{Muller11}
{M{\"u}ller-S{\'a}nchez}, F., {Prieto}, M.~A., {Hicks}, E.~K.~S., {et~al.}
  2011, \apj, 739, 69

\bibitem[{{Murray} {et~al.}(1995){Murray}, {Chiang}, {Grossman}, \&
  {Voit}}]{Murray95}
{Murray}, N., {Chiang}, J., {Grossman}, S.~A., \& {Voit}, G.~M. 1995, \apj,
  451, 498

\bibitem[{{Nagar} {et~al.}(2002){Nagar}, {Oliva}, {Marconi}, \&
  {Maiolino}}]{Nagar02}
{Nagar}, N.~M., {Oliva}, E., {Marconi}, A., \& {Maiolino}, R. 2002, \aap, 391,
  L21

\bibitem[{{Narayan}(2002)}]{Narayan02}
{Narayan}, R. 2002, in Lighthouses of the Universe: The Most Luminous Celestial
  Objects and Their Use for Cosmology, ed. {M.~Gilfanov, R.~Sunyeav, \&
  E.~Churazov}, 405

\bibitem[{{Natta} \& {Panagia}(1984)}]{Natta84}
{Natta}, A. \& {Panagia}, N. 1984, \apj, 287, 228

\bibitem[{{Netzer}(2009)}]{Netzer09}
{Netzer}, H. 2009, \apj, 695, 793

\bibitem[{{Netzer} \& {Laor}(1993)}]{NetzerLaor93}
{Netzer}, H. \& {Laor}, A. 1993, \apjl, 404, L51

\bibitem[{{Nicastro}(2000)}]{Nicastro00}
{Nicastro}, F. 2000, \apjl, 530, L65

\bibitem[{{Nussbaumer} \& {Storey}(1988)}]{nuss88}
{Nussbaumer}, H. \& {Storey}, P.~J. 1988, \aap, 193, 327

\bibitem[{{Orban de Xivry} {et~al.}(2011){Orban de Xivry}, {Davies},
  {Schartmann}, {Komossa}, {Marconi}, {Hicks}, {Engel}, \& {Tacconi}}]{Xivry11}
{Orban de Xivry}, G., {Davies}, R., {Schartmann}, M., {et~al.} 2011, \mnras,
  417, 2721

\bibitem[{{Origlia} {et~al.}(1993){Origlia}, {Moorwood}, \& {Oliva}}]{OO93}
{Origlia}, L., {Moorwood}, A.~F.~M., \& {Oliva}, E. 1993, \aap, 280, 536

\bibitem[{{Osterbrock}(1989)}]{Osterbrock89}
{Osterbrock}, D.~E. 1989, {Astrophysics of gaseous nebulae and active galactic
  nuclei} ({University Science Books})

\bibitem[{{Panessa} \& {Bassani}(2002)}]{PanessaBassani02}
{Panessa}, F. \& {Bassani}, L. 2002, \aap, 394, 435

\bibitem[{{Panessa} {et~al.}(2009){Panessa}, {Carrera}, {Bianchi}, {Corral},
  {Gastaldello}, {Barcons}, {Bassani}, {Matt}, \& {Monaco}}]{Panessa09}
{Panessa}, F., {Carrera}, F.~J., {Bianchi}, S., {et~al.} 2009, \mnras, 398,
  1951

\bibitem[{{Panessa} {et~al.}(2005){Panessa}, {Wolter}, {Pellegrini},
  {Fruscione}, {Bassani}, {Della Ceca}, {Palumbo}, \& {Trinchieri}}]{Pa05}
{Panessa}, F., {Wolter}, A., {Pellegrini}, S., {et~al.} 2005, \apj, 631, 707

\bibitem[{{Panuzzo} {et~al.}(2003){Panuzzo}, {Bressan}, {Granato}, {Silva}, \&
  {Danese}}]{Panuzzo03}
{Panuzzo}, P., {Bressan}, A., {Granato}, G.~L., {Silva}, L., \& {Danese}, L.
  2003, \aap, 409, 99

\bibitem[{Peterson {et~al.}(2005)Peterson, Bentz, Desroches, Filippenko, Ho,
  Kaspi, Laor, Maoz, Moran, Pogge, \& Quillen}]{Peterson05}
Peterson, B.~M., Bentz, M.~C., Desroches, L.-B., {et~al.} 2005, The
  Astrophysical Journal, 632, 799

\bibitem[{{Ramos Almeida} {et~al.}(2008){Ramos Almeida}, {P{\'e}rez
  Garc{\'{\i}}a}, {Acosta-Pulido}, \&
  {Gonz{\'a}lez-Mart{\'{\i}}n}}]{Ramosalmeida08}
{Ramos Almeida}, C., {P{\'e}rez Garc{\'{\i}}a}, A.~M., {Acosta-Pulido}, J.~A.,
  \& {Gonz{\'a}lez-Mart{\'{\i}}n}, O. 2008, \apjl, 680, L17

\bibitem[{{Reunanen} {et~al.}(2003){Reunanen}, {Kotilainen}, \&
  {Prieto}}]{Reunanen03}
{Reunanen}, J., {Kotilainen}, J.~K., \& {Prieto}, M.~A. 2003, \mnras, 343, 192

\bibitem[{{Rhee} \& {Larkin}(2000)}]{RheeLarkin00}
{Rhee}, J.~H. \& {Larkin}, J.~E. 2000, \apj, 538, 98

\bibitem[{{Rhee} \& {Larkin}(2005)}]{Rhee05}
{Rhee}, J.~H. \& {Larkin}, J.~E. 2005, \apj, 620, 151

\bibitem[{{Rieke} \& {Lebofsky}(1985)}]{RiekeyLebofsky85}
{Rieke}, G.~H. \& {Lebofsky}, M.~J. 1985, \apj, 288, 618

\bibitem[{{Riffel} {et~al.}(2009{\natexlab{a}}){Riffel}, {Pastoriza},
  {Rodr{\'{\i}}guez-Ardila}, \& {Bonatto}}]{Rifstarlight}
{Riffel}, R., {Pastoriza}, M.~G., {Rodr{\'{\i}}guez-Ardila}, A., \& {Bonatto},
  C. 2009{\natexlab{a}}, \mnras, 400, 273

\bibitem[{{Riffel} \& {Storchi-Bergmann}(2011{\natexlab{a}})}]{Riffel11}
{Riffel}, R.~A. \& {Storchi-Bergmann}, T. 2011{\natexlab{a}}, \mnras, 411, 469

\bibitem[{{Riffel} \& {Storchi-Bergmann}(2011{\natexlab{b}})}]{Rif11}
{Riffel}, R.~A. \& {Storchi-Bergmann}, T. 2011{\natexlab{b}},
  arXiv:astro-ph/1107.2564

\bibitem[{{Riffel} {et~al.}(2009{\natexlab{b}}){Riffel}, {Storchi-Bergmann},
  {Dors}, \& {Winge}}]{Rif09}
{Riffel}, R.~A., {Storchi-Bergmann}, T., {Dors}, O.~L., \& {Winge}, C.
  2009{\natexlab{b}}, \mnras, 393, 783

\bibitem[{{Riffel} {et~al.}(2009{\natexlab{c}}){Riffel}, {Storchi-Bergmann}, \&
  {McGregor}}]{Rif09torus}
{Riffel}, R.~A., {Storchi-Bergmann}, T., \& {McGregor}, P.~J.
  2009{\natexlab{c}}, \apj, 698, 1767

\bibitem[{{Riffel} {et~al.}(2010){Riffel}, {Storchi-Bergmann}, \&
  {Nagar}}]{Rif10}
{Riffel}, R.~A., {Storchi-Bergmann}, T., \& {Nagar}, N.~M. 2010, \mnras, 404,
  166

\bibitem[{{Risaliti} {et~al.}(2005){Risaliti}, {Elvis}, {Fabbiano}, {Baldi}, \&
  {Zezas}}]{Risaliti05}
{Risaliti}, G., {Elvis}, M., {Fabbiano}, G., {Baldi}, A., \& {Zezas}, A. 2005,
  \apjl, 623, L93

\bibitem[{{Rodr{\'{\i}}guez-Ardila}
  {et~al.}(2005{\natexlab{a}}){Rodr{\'{\i}}guez-Ardila}, {Contini}, \&
  {Viegas}}]{ardila05nlsy}
{Rodr{\'{\i}}guez-Ardila}, A., {Contini}, M., \& {Viegas}, S.~M.
  2005{\natexlab{a}}, \mnras, 357, 220

\bibitem[{{Rodr{\'{\i}}guez-Ardila} {et~al.}(2004){Rodr{\'{\i}}guez-Ardila},
  {Pastoriza}, {Viegas}, {Sigut}, \& {Pradhan}}]{ardila04}
{Rodr{\'{\i}}guez-Ardila}, A., {Pastoriza}, M.~G., {Viegas}, S., {Sigut},
  T.~A.~A., \& {Pradhan}, A.~K. 2004, \aap, 425, 457

\bibitem[{{Rodr{\'{\i}}guez-Ardila}
  {et~al.}(2005{\natexlab{b}}){Rodr{\'{\i}}guez-Ardila}, {Riffel}, \&
  {Pastoriza}}]{ardila05}
{Rodr{\'{\i}}guez-Ardila}, A., {Riffel}, R., \& {Pastoriza}, M.~G.
  2005{\natexlab{b}}, \mnras, 364, 1041

\bibitem[{{Salpeter}(1977)}]{Salpeter77}
{Salpeter}, E.~E. 1977, \araa, 15, 267

\bibitem[{{Schlegel} {et~al.}(1998){Schlegel}, {Finkbeiner}, \&
  {Davis}}]{Schlegel98}
{Schlegel}, D.~J., {Finkbeiner}, D.~P., \& {Davis}, M. 1998, \apj, 500, 525

\bibitem[{{Shemmer} {et~al.}(2008){Shemmer}, {Brandt}, {Netzer}, {Maiolino}, \&
  {Kaspi}}]{Shemmer08}
{Shemmer}, O., {Brandt}, W.~N., {Netzer}, H., {Maiolino}, R., \& {Kaspi}, S.
  2008, \apj, 682, 81

\bibitem[{{Shen} {et~al.}(2010){Shen}, {Shao}, \& {Gu}}]{Shen10}
{Shen}, S., {Shao}, Z., \& {Gu}, M. 2010, \apjl, 725, L210

\bibitem[{{Shi} {et~al.}(2010){Shi}, {Rieke}, {Smith}, {Rigby}, {Hines},
  {Donley}, {Schmidt}, \& {Diamond-Stanic}}]{Shi10}
{Shi}, Y., {Rieke}, G.~H., {Smith}, P., {et~al.} 2010, \apj, 714, 115

\bibitem[{{Singh} {et~al.}(2011){Singh}, {Shastri}, \& {Risaliti}}]{Singh11}
{Singh}, V., {Shastri}, P., \& {Risaliti}, G. 2011, \aap, 533, A128

\bibitem[{{Skrutskie} {et~al.}(2006){Skrutskie}, {Cutri}, {Stiening},
  {Weinberg}, {Schneider}, {Carpenter}, {Beichman}, {Capps}, {Chester},
  {Elias}, {Huchra}, {Liebert}, {Lonsdale}, {Monet}, {Price}, {Seitzer},
  {Jarrett}, {Kirkpatrick}, {Gizis}, {Howard}, {Evans}, {Fowler}, {Fullmer},
  {Hurt}, {Light}, {Kopan}, {Marsh}, {McCallon}, {Tam}, {Van Dyk}, \&
  {Wheelock}}]{2mass}
{Skrutskie}, M.~F., {Cutri}, R.~M., {Stiening}, R., {et~al.} 2006, \aj, 131,
  1163

\bibitem[{{Stern} \& {Laor}(2012)}]{SternLaor2012}
{Stern}, J. \& {Laor}, A. 2012, \mnras, 2870

\bibitem[{{Sternberg}(1998)}]{Sternberg98}
{Sternberg}, A. 1998, \apj, 506, 721

\bibitem[{{Storchi-Bergmann} {et~al.}(2009){Storchi-Bergmann}, {McGregor},
  {Riffel}, {Sim{\~o}es Lopes}, {Beck}, \& {Dopita}}]{Sch09}
{Storchi-Bergmann}, T., {McGregor}, P.~J., {Riffel}, R.~A., {et~al.} 2009,
  \mnras, 394, 1148

\bibitem[{{Terashima} {et~al.}(2000){Terashima}, {Ho}, \& {Ptak}}]{Terashima00}
{Terashima}, Y., {Ho}, L.~C., \& {Ptak}, A.~F. 2000, \apj, 539, 161

\bibitem[{{Thornley} {et~al.}(2000){Thornley}, {Schreiber}, {Lutz}, {Genzel},
  {Spoon}, {Kunze}, \& {Sternberg}}]{Thornley00}
{Thornley}, M.~D., {Schreiber}, N.~M.~F., {Lutz}, D., {et~al.} 2000, \apj, 539,
  641

\bibitem[{{Thornton} {et~al.}(2009){Thornton}, {Barth}, {Ho}, \&
  {Greene}}]{Thornton09}
{Thornton}, C.~E., {Barth}, A.~J., {Ho}, L.~C., \& {Greene}, J.~E. 2009, \apj,
  705, 1196

\bibitem[{Thornton {et~al.}(2008)Thornton, Barth, Ho, Rutledge, \&
  Greene}]{Thornton08}
Thornton, C.~E., Barth, A.~J., Ho, L.~C., Rutledge, R.~E., \& Greene, J.~E.
  2008, \apj, 686, 892

\bibitem[{{Tokunaga} \& {Vacca}(2005)}]{TokunagaandVacca05}
{Tokunaga}, A.~T. \& {Vacca}, W.~D. 2005, \pasp, 117, 421

\bibitem[{{Tran}(2001)}]{Tran01}
{Tran}, H.~D. 2001, \apjl, 554, L19

\bibitem[{{Tran} {et~al.}(2011){Tran}, {Lyke}, \& {Mader}}]{Tran11}
{Tran}, H.~D., {Lyke}, J.~E., \& {Mader}, J.~A. 2011, \apjl, 726, L21

\bibitem[{{Tremaine} {et~al.}(2002){Tremaine}, {Gebhardt}, {Bender}, {Bower},
  {Dressler}, {Faber}, {Filippenko}, {Green}, {Grillmair}, {Ho}, {Kormendy},
  {Lauer}, {Magorrian}, {Pinkney}, \& {Richstone}}]{Tremaine02}
{Tremaine}, S., {Gebhardt}, K., {Bender}, R., {et~al.} 2002, \apj, 574, 740

\bibitem[{{Trump} {et~al.}(2011){Trump}, {Impey}, {Kelly}, {Civano}, {Gabor},
  {Diamond-Stanic}, {Merloni}, {Urry}, {Hao}, {Jahnke}, {Nagao}, {Taniguchi},
  {Koekemoer}, {Lanzuisi}, {Liu}, {Mainieri}, {Salvato}, \&
  {Scoville}}]{Trump11}
{Trump}, J.~R., {Impey}, C.~D., {Kelly}, B.~C., {et~al.} 2011, \apj, 733, 60

\bibitem[{{Urry} \& {Padovani}(1995)}]{Urry95}
{Urry}, C.~M. \& {Padovani}, P. 1995, \pasp, 107, 803

\bibitem[{{Valencia-S.} {et~al.}(2012){Valencia-S.}, {Eckart}, {Zuther},
  {Fischer}, {Smajic}, {Iserlohe}, {Garc\'{\i}a-Mar\'{\i}n}, {Moser}, {Bremer},
  \& {Vitale}}]{me2012a}
{Valencia-S.}, M., {Eckart}, A., {Zuther}, J., {et~al.} 2012, in {J}. Phys.:
  Conf. Ser., (accepted)

\bibitem[{{Valencia-S.} \& {et al.}(2012)}]{me2012b}
{Valencia-S.}, M. \& {et al.} 2012, (in preparation)

\bibitem[{{Vasudevan} \& {Fabian}(2007)}]{vasudevan07}
{Vasudevan}, R.~V. \& {Fabian}, A.~C. 2007, \mnras, 381, 1235

\bibitem[{{V{\'a}zquez} \& {Leitherer}(2005)}]{Vazquez05}
{V{\'a}zquez}, G.~A. \& {Leitherer}, C. 2005, \apj, 621, 695

\bibitem[{{Veilleux} {et~al.}(1997){Veilleux}, {Goodrich}, \&
  {Hill}}]{Veilleux97}
{Veilleux}, S., {Goodrich}, R.~W., \& {Hill}, G.~J. 1997, \apj, 477, 631

\bibitem[{{Veilleux} \& {Osterbrock}(1987)}]{VO87}
{Veilleux}, S. \& {Osterbrock}, D.~E. 1987, \apjs, 63, 295

\bibitem[{{Ward} {et~al.}(1988){Ward}, {Done}, {Fabian}, {Tennant}, \&
  {Shafer}}]{Ward88}
{Ward}, M.~J., {Done}, C., {Fabian}, A.~C., {Tennant}, A.~F., \& {Shafer},
  R.~A. 1988, \apj, 324, 767

\bibitem[{{Willner} {et~al.}(1984){Willner}, {Fabbiano}, {Elvis}, {Ward},
  {Longmore}, \& {Lawrence}}]{willner84}
{Willner}, S.~P., {Fabbiano}, G., {Elvis}, M., {et~al.} 1984, \pasp, 96, 143

\bibitem[{{Winge} {et~al.}(2009){Winge}, {Riffel}, \&
  {Storchi-Bergmann}}]{Winge09}
{Winge}, C., {Riffel}, R.~A., \& {Storchi-Bergmann}, T. 2009, \apjs, 185, 186

\bibitem[{{Woo} {et~al.}(2010){Woo}, {Treu}, {Barth}, {Wright}, {Walsh},
  {Bentz}, {Martini}, {Bennert}, {Canalizo}, {Filippenko}, {Gates}, {Greene},
  {Li}, {Malkan}, {Stern}, \& {Minezaki}}]{Woo10}
{Woo}, J.-H., {Treu}, T., {Barth}, A.~J., {et~al.} 2010, \apj, 716, 269

\bibitem[{{Xiao} {et~al.}(2011){Xiao}, {Barth}, {Greene}, {Ho}, {Bentz},
  {Ludwig}, \& {Jiang}}]{Xiao11}
{Xiao}, T., {Barth}, A.~J., {Greene}, J.~E., {et~al.} 2011, \apj, 739, 28

\bibitem[{{Zhang} \& {Wang}(2006)}]{Zhang06}
{Zhang}, E.-P. \& {Wang}, J.-M. 2006, \apj, 653, 137

\bibitem[{Zhou {et~al.}(2006)Zhou, Wang, Yuan, Lu, Dong, Wang, \& Lu}]{Zhou06}
Zhou, H., Wang, T., Yuan, W., {et~al.} 2006, \apjs, 166, 128

\bibitem[{{Zuther} {et~al.}(2007){Zuther}, {Iserlohe}, {Pott}, {Bertram},
  {Fischer}, {Voges}, {Hasinger}, \& {Eckart}}]{Zuther}
{Zuther}, J., {Iserlohe}, C., {Pott}, J.-U., {et~al.} 2007, \aap, 466, 451

\end{thebibliography}

\end{document}